\documentclass[11pt]{article}

\usepackage{verbatim}
\usepackage{amsmath}
\usepackage{amsfonts}
\usepackage{amssymb}
\usepackage{bbm}
\usepackage{latexsym}
\usepackage{lscape} 
\usepackage{epsfig}
\usepackage{color}
\usepackage{graphicx}
\usepackage[table]{xcolor}
\usepackage{enumerate}
\usepackage{hhline}
\usepackage{subfig}
\usepackage{multirow}
\usepackage{mathrsfs}
\usepackage{longtable}
\usepackage{hhline}

\usepackage[titletoc,title]{appendix}
\usepackage{amscd}
\usepackage{cite}

\renewcommand{\arraystretch}{1.5} 

\DeclareMathSymbol{\mg}{\mathrel}{symbols}{"1D}

\newcommand{\ga}{\alpha}
\newcommand{\gb}{\beta}
\renewcommand{\gg}{\gamma}
\newcommand{\gd}{\delta}

\newcommand{\gvf}{\varphi}

\newcommand{\gm}{\mu}
\newcommand{\gn}{\nu}

\newcommand{\gk}{\kappa}
\newcommand{\gl}{\lambda}

\newcommand{\gth}{\theta}

\newcommand{\gs}{\sigma}

\newcommand{\go}{\omega}

\newcommand{\gp}{\pi}
\newcommand{\gps}{\psi}

\newcommand{\gG}{\Gamma}
\newcommand{\gD}{\Delta}
\newcommand{\gF}{\Phi}

\newcommand{\gL}{\Lambda}

\newcommand{\gO}{\Omega}

\newcommand{\gPs}{\Psi}

\newcommand{\cA}{{\cal A}}

\newcommand{\cD}{{\cal D}}

\newcommand{\cF}{{\cal F}}

\newcommand{\cJ}{{\cal J}}
\newcommand{\cK}{{\cal K}}

\newcommand{\cR}{{\cal R}}

\newcommand{\cV}{{\cal V}}
\newcommand{\cW}{{\cal W}}

\newcommand{\ua}{{\underline a}}

\newcommand{\um}{{\underline m}}

\newcommand{\ta}{{\tilde a}}
\newcommand{\tb}{{\tilde b}}

\newcommand{\tn}{{\tilde n}}

\newcommand{\tF}{{\widetilde F}}

\newcommand{\tN}{{\widetilde N}}

\newcommand{\dga}{{\dot \alpha}}

\newcommand{\Tr}{\mbox{Tr}}
\newcommand{\tr}{\text{tr}}

\newcommand{\Id}{\text{\small 1}\hspace{-3.5pt}\text{1}}

\newcommand{\ra}{\rightarrow}

\newcommand{\der}{\partial}

\newcommand{\dsp}{\displaystyle}

\newcommand{\undr}[1]{{\underline{#1}}}

\newcommand{\beq}{\begin{equation}}
\newcommand{\eeq}{\end{equation}}
\newcommand{\barr}{\begin{array}}
\newcommand{\earr}{\end{array}}
\newcommand{\equ}[1]{\begin{gather} #1 \end{gather}}
\newcommand{\equa}[1]{\begin{align} #1 \end{align}}
\newcommand{\items}[1]{\begin{itemize} #1 \end{itemize}}
\newcommand{\enums}[1]{\begin{enumerate} #1 \end{enumerate}}

\newcommand{\arry}[2]{\begin{array}{#1} #2 \end{array}}

\newcommand{\pmtrx}[1]{\begin{pmatrix} #1 \end{pmatrix}}

\newcommand{\non}{\nonumber}
\newcommand{\sfrac}[2]{\mbox{$\frac{#1}{#2}$}}
\newcounter{oldcounter}

\newcommand{\bD}{{\overline D}}

\newcommand{\bZ}{{\overline Z}}
\newcommand{\bgth}{{\bar\theta}}

\newcommand{\bgO}{{\overline\Omega}}

\newcommand{\Intr}{\mathbb{Z}}

\newcommand{\ba}[2]{\[\begin{array}{#2}\label{#1}}
\newcommand{\ea}{\end{array}\]}
\newcommand{\be}{\begin{equation}}
\newcommand{\ee}{\end{equation}}
\newcommand{\bea}{\begin{eqnarray}}
\newcommand{\eea}{\end{eqnarray}}

\newcommand{\rep}[1]{\mathbf{#1}}
\newcommand{\crep}[1]{\overline{\rep{#1}}}
\newcommand{\brep}[1]{\overline{\rep{#1}}}

\newcommand{\sm}{{\,\mbox{-}}}

\usepackage{cancel}
\usepackage{hyperref}

\begin{document}

\thispagestyle{empty}

\begin{flushright}
DESY-15-125, LMU-ASC 46/15, MITP/15-053\\
\end{flushright}

\begin{center}
{\Large {\bf 
Calabi-Yau compactifications of non-supersymmetric heterotic string theory
} 
}
\\[0pt]

\bigskip
\bigskip 
{\large
{\bf{Michael Blaszczyk}$^{a,}$\footnote{E-mail: blaszcz@uni-mainz.de}},
{\bf{Stefan Groot Nibbelink}$^{b,}$\footnote{E-mail: Groot.Nibbelink@physik.uni-muenchen.de}},  
{\bf{Orestis Loukas}$^{b,c,}$\footnote{E-mail: O.Loukas@physik.uni-muenchen.de}},\\[1ex]
{\bf{Fabian Ruehle$^{d,}$}\footnote{E-mail: fabian.ruehle@desy.de}}
\bigskip}\\[0pt]
\vspace{0.23cm}
${}^a$ {\it 
PRISMA Cluster of Excellence \& Institut f\"ur Physik (WA THEP), \\ 
Johannes-Gutenberg-Universit\"at, 55099 Mainz, Germany
}\\[1ex] 
${}^b$ {\it 
Arnold Sommerfeld Center for Theoretical Physics,   \\ 
Ludwig-Maximilians-Universit\"at M\"unchen, 80333 M\"unchen, Germany
}
\\[1ex]
${}^c$ {\it School of Electrical and Computer Engineering, \\
National Technical University of Athens, Zografou Campus, GR-15780 Athens, Greece}
\\[1ex]
${}^d$ {\it 
Deutsches Elektronen-Synchrotron DESY, Notkestrasse 85, 22607 Hamburg, Germany
}
\\[10ex] 
\end{center}

\subsection*{\centering Abstract}
Phenomenological explorations of heterotic strings have conventionally focused primarily on the E$_8\times$E$_8$ theory. We consider smooth compactifications of all three ten-dimensional heterotic theories to exhibit the many similarities between the non-supersymmetric SO(16)$\times$SO(16) theory and the related supersymmetric E$_8\times$E$_8$ and SO(32) theories. In particular, we exploit these similarities to determine the bosonic and fermionic spectra of Calabi-Yau compactifications with line bundles of the non-supersymmetric string. We use elements of four-dimensional supersymmetric effective field theory to characterize the non-supersymmetric action at leading order and determine the Green-Schwarz induced axion-couplings. Using these methods we construct a non-supersymmetric Standard Model(SM)-like theory. In addition, we show that it is possible to obtain SM-like models from the standard embedding using at least an order four Wilson line. Finally, we make a proposal of the states that live on five branes in the SO(16)$\times$SO(16) theory and find under certain assumptions the surprising result that anomaly factorization only admits at most a single brane solution.

\newpage 
\setcounter{page}{1}
\setcounter{footnote}{0}
\tableofcontents
\newpage

\section{Introduction and summary}
\label{sc:Introduction}

One of the most frequently considered extensions of the Standard Model (SM) of Particle Physics is supersymmetry. This hypothetical symmetry assigns to each observed particle a supersymmetric partner which has identical properties except that their spins differ by 1/2, i.e.\ it relates bosons to fermions and vice versa. One of the main goals of the LHC accelerator at CERN is to observe these superpartners. So far, there have been no hints for the existence of such states. 

One may wonder about the consequences for string theory, if supersymmetry will not be found at the LHC or possible future accelerators. In light of the often stated claim that string theory predicts supersymmetry, this seems to be a doomsday scenario for string theory. However, the statement that string theory requires target space supersymmetry is simply false: As had been realized, essentially during the time that string theory was first considered as a unified framework of all particles and interactions, it is possible to construct consistent string theories without space-time supersymmetry. Minimal requirements on a consistent theory are modular invariance and the absence of anomalies and tachyons. A prime example of a non-supersymmetric string theory is the SO(16)$\times$SO(16) string~\cite{Dixon:1986iz,Dixon:1986jc,AlvarezGaume:1986jb}.  

In the past decades various authors have considered non-supersymmetric string constructions. 
Torus compactifications with Wilson lines of the SO(16)$\times$SO(16) theory have been studied in Ref.~\cite{Nair:1986zn,Ginsparg:1986wr}. 
In addition, using a covariant lattice approach four-dimensional non-supersymmetric models were constructed by~\cite{Lerche:1986ae,Lerche:1986cx}.
Compactification on (a)symmetric orbifolds~\cite{Dixon:1985jw,Dixon:1986jc,Ibanez:1986tp,Ibanez:1987pj} of the non-supersymmetric SO(16)$\times$SO(16) have been inspected in the papers~\cite{Taylor:1987uv,Toon:1990ij,Sasada:1995wq,Font:2002pq}. 
Non-supersymmetric compactifications of the heterotic string have also been investigated using the free-fermionic string description~~\cite{Kawai:1986vd,Antoniadis:1986rn,Antoniadis:1987wp} see e.g.\ the works~\cite{Dienes:1994np,Blum:1997cs,Shiu:1998he,Dienes:2006ut,Faraggi:2007tj}. 
Also in non-heterotic string context tachyon-free non-supersymmetric models have been constructed, for example as type-II orientifolds~\cite{Sagnotti:1995ga,Sagnotti:1996qj,Angelantonj:1998gj,Sugimoto:1999tx,Blumenhagen:1999ns, Aldazabal:1999tw,Moriyama:2001ge} or rational conformal field theories~\cite{GatoRivera:2007yi,GatoRivera:2008zn}.

Last year various new investigations of the phenomenological potential of non-supersymmetric compactifications of the heterotic string have appeared. In~\cite{Blaszczyk:2014qoa,Nibbelink:2015ena} some of the authors of the current paper considered orbifold and smooth Calabi-Yau compactifications of the non-supersymmetric SO(16)$\times$SO(16) string. It was argued that Calabi-Yau compactification of the SO(16)$\times$SO(16) theory has particular phenomenological potential, since tachyons can be avoided to leading orders in the $\ga'$ and $g_s$ expansions. Moreover, it was shown that it is possible to obtain many tachyon-free orbifold models with spectra quite close to the Standard Model (SM). 

Also certain half-flat geometries can be considered as non-supersymmetric backgrounds for the heterotic string~\cite{Lukas:2015kca}. In addition, semi-realistic models were constructed in the free-fermionic context in~\cite{Abel:2015oxa,Ashfaque:2015vta}, which implement the idea of having models that  interpolate between supersymmetric and non-supersymmetric string constructions
~\cite{Itoyama:1986ei,Dienes:1990ij,Dienes:1994np,Faraggi:2009xy}. For such models it is possible to compute more detailed phenomenological quantities like threshold corrections~\cite{Angelantonj:2014dia,Florakis:2015txa}. 

\begin{figure}[t!]
\centering\includegraphics[width=.8\textwidth]{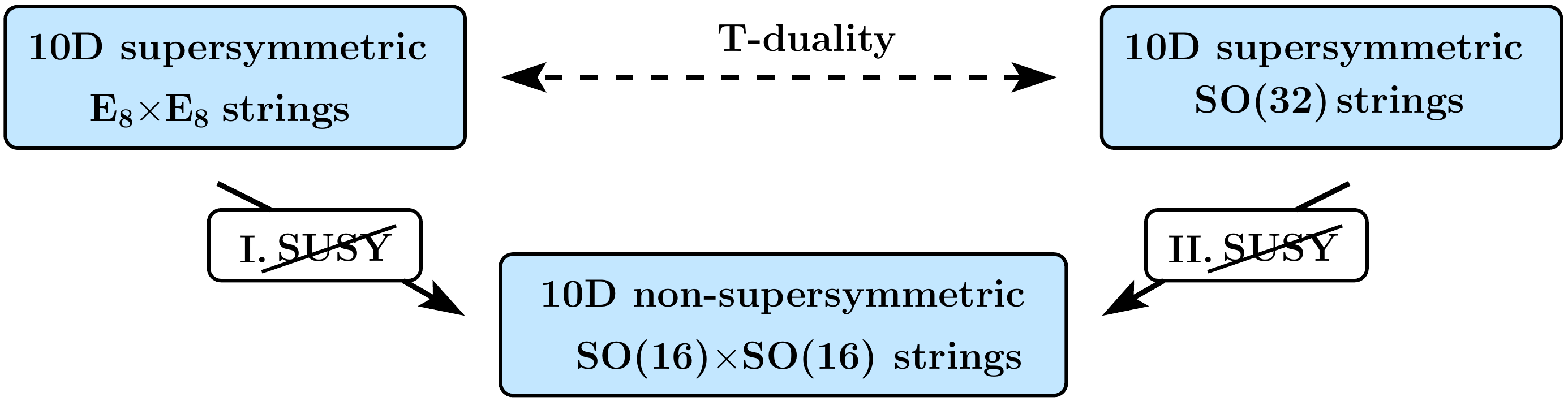}
\caption{\label{fg:RelationHetStrings}
This figure depicts the relation between the three heterotic string theories in ten dimensions.} 
\end{figure}

In this paper we lay out more theoretical methods to perform phenomenological investigations of smooth compactifications of the non-supersymmetric SO(16)$\times$SO(16) string. In order to do so, we often make use of the fact that the non-supersymmetric heterotic SO(16)$\times$SO(16) string is closely related to the supersymmetric heterotic E$_8\times$E$_8$ and SO(32) strings: It is well-known that upon compactification on a circle with appropriate Wilson lines both supersymmetric heterotic strings become T-dual to each other. Moreover, the non-supersymmetric SO(16)$\times$SO(16) theory can be obtained by supersymmetry-breaking twists acting on either the E$_8\times$E$_8$ or the SO(32) theories. These relations between the three heterotic theories in ten dimensions are indicated in Figure~\ref{fg:RelationHetStrings}. Interestingly, the full low energy spectrum of the SO(16)$\times$SO(16) theory can be obtained by simple orbifold projections and interpreted as the combined untwisted sectors of the E$_8\times$E$_8$ and the SO(32) strings. As can be inferred from Table~\ref{tb:10DhetSpectra}, the gravitational and gauge sectors can be obtained from either route. On the other hand, the chiral fermions in the spinor representation of the SO(16) factors come from the untwisted sector of the E$_8\times$E$_8$ string, while the chiral fermions in the bi-fundamental of SO(16)$\times$SO(16) are part of its twisted sector w.r.t.\ the supersymmetry-breaking twist. In contrast, for the SO(32) theory the roles of untwisted and twisted chiral matter are precisely interchanged.

Next we investigate the effective four-dimensional theories that arise when we compactify one of the heterotic strings on Calabi-Yau geometries with (line) bundles. When we start from the non-supersymmetric SO(16)$\times$SO(16) theory, there seems to be no need to consider string backgrounds that would themselves preserve some amount of supersymmetry. However, there are various reasons why insisting on Calabi-Yau geometries with holomorphic stable vector bundles is convenient: 

From a worldsheet point of view, having a complex manifold with a holomorphic vector bundle means that one has an enhanced global U(1)$_R$ symmetry so that the worldsheet theory has (2,0) supersymmetry. This U(1)$_R$ symmetry is non-anomalous precisely if the manifold is Calabi-Yau, see~\cite{Hull:1985jv,Hull:1990qh}. Notice that, to leading order, these arguments do not depend on the global boundary conditions on the worldsheet, i.e.\ the spin-structures and hence they apply to each of the three heterotic theories. 

Also from the target space Effective Field Theory (EFT) perspective, there are good reasons to consider supersymmetry-preserving compactifications of the SO(16)$\times$SO(16) string. As stated above, it was shown in~\cite{Blaszczyk:2014qoa,Nibbelink:2015ena} that such compactifications avoid tachyons to leading order in $g_s$ and $\ga'$. Moreover, the bosonic parts of the three heterotic ten-dimensional theories are identical up to their gauge groups. Consequently, the reduction of either of these theories on the same background leads to essentially identical EFTs in four dimensions. This means that the effective potential of SO(16)$\times$SO(16) compactifications, relevant to characterize the physical vacuum, is characterized by the same quantities as compactifications of its supersymmetric cousins, at least to leading order. This shows that stable supersymmetric backgrounds (solutions to F- and D-term conditions at tree-level) also represent solutions to the field equations of the non-supersymmetric SO(16)$\times$SO(16) theory. Also it turns out to be fruitful to employ concepts like the super- and K\"ahler potential to characterize its four-dimensional EFT.

\begin{table}
\[
\begin{array}{|p{4cm}||c|c|c|} 
\hline 
\centering\text{\bf Theory} & \text{\bf Sector} & \cellcolor{lightgray}   \text{\bf Bosons} & \text{\bf Fermions}  
\\ \hline\hline 
\multirow{2}{4cm}{\centering\text{\bf supersymmetric}\newline\text{\bf E$\boldsymbol{_8\times}$E$\boldsymbol{_8}$}} & \text{gravity} & \cellcolor{lightgray}  \text{metric, B-field, dilaton} & \text{gravitinos, dilatinos}  
\\ \hhline{|~||-|-|-|}
& \text{gauge} & \cellcolor{lightgray} (\mathbf{248},\mathbf{1})+(\mathbf{1},\mathbf{248}) \text{ gauge fields} & (\mathbf{248},\mathbf{1})+(\mathbf{1},\mathbf{248}) \text{ gauginos}
\\ \hline\hline 
\multirow{2}{4cm}{\centering\text{\bf supersymmetric}\newline\text{\bf SO(32)}} & \text{gravity} & \cellcolor{lightgray} \text{metric, B-field, dilaton} & \text{gravitinos, dilatinos}  
\\ \hhline{|~||-|-|-|}
& \text{gauge} & \cellcolor{lightgray}  \mathbf{496} \text{ gauge fields} & \mathbf{496} \text{ gauginos}
\\ \hline\hline 
\multirow{4}{4cm}{\centering{\rm \bf non-supersymmetric SO(16)$\boldsymbol{\times}$SO(16)}} &  \text{gravity} &  \cellcolor{lightgray} \text{metric, B-field, dilaton} & \text{}  
\\ \hhline{~||-|-|~}
& \text{gauge} & \cellcolor{lightgray}  (\mathbf{120},\mathbf{1})+(\mathbf{1},\mathbf{120}) \text{ gauge fields} & \text{}
\\ \hhline{|~||-|-|-|}
& \multirow{2}{*}{matter} & \cellcolor{lightgray} & (\mathbf{128},\mathbf{1}) + (\mathbf{1},\mathbf{128}) \text{ spinors} \\
&&\cellcolor{lightgray}& (\mathbf{16},\mathbf{16}) \text{ co-spinors}   
\\ \hline 
\end{array} 
\]
\caption{This table gives the bosonic and fermionic spectra of the three consistent heterotic string theories with gauge groups E$_8\times$E$_8$, SO(32) and SO(16)$\times$SO(16).\label{tb:10DhetSpectra}}
\end{table}

All this suggests that many methods developed for Calabi-Yau compactifications can be used to obtain results for compactifications of the non-supersymmetric SO(16)$\times$SO(16) theory as well. The supersymmetric heterotic theories~\cite{Blumenhagen:2005ga,Blumenhagen:2005pm,Blumenhagen:2006ux}, and in particular the E$_8\times$E$_8$ theory  with non-Abelian bundles~\cite{Braun:2005ux,Bouchard:2005ag,Anderson:2007nc} or line bundles on orbifold resolutions~\cite{Nibbelink:2007rd,Nibbelink:2009sp,Blaszczyk:2010db} and Complete Intersection Calabi-Yau manifolds (CICYs)~\cite{Anderson:2011ns,Anderson:2012yf}, have been well-studied since the seminal paper~\cite{Candelas:1985en}. (A technical side result that we derive in Appendix~\ref{sc:IdUbundles} is how to translate the line bundle parameterization used in~\cite{Anderson:2011ns,Anderson:2012yf} to the line bundle vector language of~\cite{Nibbelink:2007rd,Nibbelink:2009sp,Blaszczyk:2010db} following~\cite{Strominger:1986uh}.) For the non-supersymmetric SO(16)$\times$SO(16) there are in principle three ways to get access to the effective theory in four dimensions as depicted in Figure~\ref{fg:CompactHetStrings}: 

\begin{figure}[t!]
\centering
\includegraphics[width=.8\textwidth]{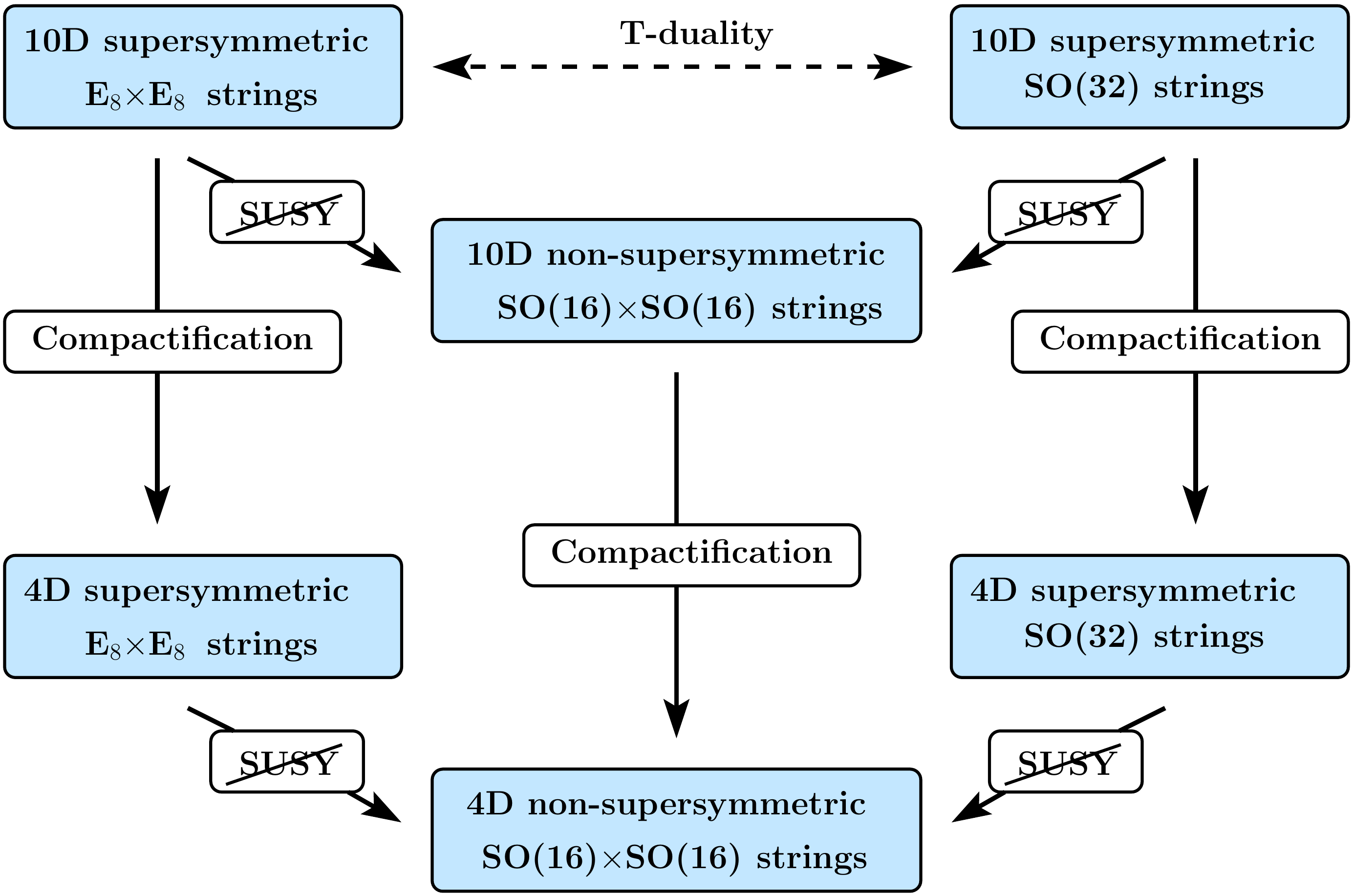}
\caption{\label{fg:CompactHetStrings}
This double commutative diagram sketches the different routes that can be taken to determine the four-dimensional effective theory by compactification of the SO(16)$\times$SO(16) theory on Calabi-Yau manifolds with holomorphic (line) bundles.} 
\end{figure}

The most direct route is indicated in the middle: One starts with the effective ten-dimensional non-supersymmetric SO(16)$\times$SO(16) theory and compactifies it on a supersymmetry-preserving background. However, given that the supersymmetry-breaking twists do not act on the internal geometry at all, we can alternatively first compactify either of the supersymmetric theories on the same smooth background, and subsequently apply the supersymmetry-breaking twists. In other words Figure~\ref{fg:CompactHetStrings} displays three alternative routes to obtain the compactifications of the SO(16)$\times$SO(16) theory. 
This means that the spectrum of gauge fields and charged scalars obtained in Calabi-Yau compactifications of the SO(16)$\times$SO(16) can be verified by compactifying either the E$_8\times$E$_8$ or SO(32) theory on the same background and subsequently applying the supersymmetry-breaking projections. Since the fermions of the SO(16)$\times$SO(16) theory are both twisted and untwisted w.r.t.\ either supersymmetric string, one needs both the E$_8\times$E$_8$ and SO(32) compactification routes to determine the full charged chiral spectrum in four dimensions. As a consequence, many properties of compactifications of the non-supersymmetric SO(16)$\times$SO(16) theory are closely related to results obtained in the past for compactifications of both supersymmetric heterotic strings. 

This applies in particular to the Green-Schwarz anomaly cancellation mechanism. It is well-known that the Green-Schwarz mechanism is very important to obtain consistent string constructions both in ten and four dimensions since it ensures the cancellation of reducible anomalies. Moreover, it determines the couplings between model-dependent and independent axions and the gauge fields. For the Calabi-Yau compactifications of the supersymmetric heterotic theories these couplings were worked out in detail in~
\cite{Blumenhagen:2005ga,Blumenhagen:2005pm,Blumenhagen:2006ux}. In this paper we investigate the Green-Schwarz mechanism for smooth Calabi-Yau compactifications of the SO(16)$\times$SO(16) string with line bundles. Using this we can perform model independent checks of anomaly cancellation for the chiral fermionic four-dimensional spectra obtained from these compactifications. 

In this work we not only want to lay out the general framework of smooth compactifications of the non-supersymmetric SO(16)$\times$SO(16) string, but we also want to show that it has true SM-like model building potential, as has been recently established for orbifold compactifications and free-fermionic constructions, as mentioned above. 
We argue that it is possible to obtain SM-like models from the standard embedding if the Calabi-Yau geometry admits at least an order four Wilson line. 
In addition, we present a particular SM-like model obtained on the smooth CICY with number 7862 (sometimes referred to as the tetra quadric) in the database~\cite{Candelas:1987kf,CYweb,Braun:2010vc}
with line bundles. We construct a six generation non-supersymmetric SU(5) GUT, which upon using a freely acting Wilson line becomes a three generation SM-like model. In a follow-up work we will present some extensive model scans on various smooth geometries and the search for SM-like models. 
However, from our current analysis we can make one interesting observation concerning the Higgs sector of such compactifications: As we will explain in this paper, we can either have a single Higgs doublet together with a color triplet partner or we have at least one pair of vector-like Higgs-doublets. 

Phenomenological model building on smooth compactifications of the E$_8\times$E$_8$ supergravity often makes use of the possibility of having five-branes. NS5-branes give an additional degree of freedom in model constructions since one does not have to satisfy the Bianchi identities strictly but only modulo effective curve classes. Effectiveness of these curves is crucial in order to guarantee that the same type of supersymmetry is preserved by the perturbative compactification and the non-perturbative NS5-brane sector. Contrary to the E$_8\times$E$_8$ theory, NS5-branes in the SO(32) context modify the spectrum charged under the perturbative gauge group. As far as we are aware five-branes in the non-supersymmetric SO(16)$\times$SO(16) context have not been studied systematically in the literature. To gain some intuition for the possible properties of NS5-branes we take inspiration from the diagram in Figure~\ref{fg:CompactHetStrings}. 

Since the fermions are the prime contributors to the anomalies, and since it is known which kind of NS5-branes the two supersymmetric heterotic theories require, we make an educated guess for the spectra on five-branes in the SO(16)$\times$SO(16) theory: We take the spectra on the NS5-branes of the E$_8\times$E$_8$ or SO(32) theories and extend the supersymmetry-breaking twist to them. We find that this choice is essentially unique if we require that all irreducible gauge and gravitational anomalies cancel. If we want in addition to cancel all reducible anomalies via a generalized Green-Schwarz mechanism, e.g.\ involving scalars and tensors on the five-branes, we find (within the ansatz that we made for these couplings) that we can only achieve this with no NS5-branes or with a configuration with one E$_8\times$E$_8$-like and one SO(32)-like NS5-brane. 

\subsection*{Acknowledgments}

We thank Steve Abel, Keith Dienes and Saul Ramos-Sanchez for valuable discussions and correspondence concerning non-supersymmetric model building. 
In addition, we would like to thank Ralph Blumenhagen and Anamaria Font for useful discussions on five-branes and anomaly cancellation in the SO(16)$\times$SO(16) context. 
Finally, we are especially indebted to Patrick K.S.~Vaudrevange for numerous discussions comparing supersymmetric and non-supersymmetric model building.

MB acknowledges the support of the {\it Cluster of Excellence `Precision Physics, Fundamental Interactions and Structure of Matter' (PRISMA)} DGF no. EXC 1098.
The work of F.R.\ was supported by the German Science Foundation (DFG) within the Collaborative Research Center (SFB) 676 ``Particles, Strings and the Early Universe''. 
O.L. acknowledges the support by the DAAD Scholarship Programme ``Vollstipendium f\"{u}r Absolventen von deutschen Auslandsschulen'' within the ``PASCH--Initiative''.

\section{Ten-dimensional heterotic strings}
\label{sc:10Dheterotic}

Conventionally, consistent string theories are characterized as string constructions that have low energy spectra which are free of anomalies and tachyons and have a modular invariant one-loop partition function. In ten dimensions there are three consistent heterotic string theories in this sense: The two that are most commonly studied, the E$_8\times$E$_8$ and SO(32) theories, are supersymmetric in target space. The third theory with gauge group SO(16)$\times$SO(16) is non-supersymmetric. It is common to distinguish the three heterotic theories by their ten-dimensional gauge group, as will be done here as well.

\subsection{Effective ten-dimensional heterotic actions}

Since we only concentrate on smooth compactifications, it is for most purposes sufficient to only consider the massless bosonic and fermionic spectrum in ten dimensions, which we give in Table~\ref{tb:10DhetSpectra} for the three heterotic theories. Their effective target space descriptions are very similar. In the string frame their bosonic action is given by 
\equ{  \label{10Daction} 
S_{10D} = \frac 1{2 \gk_{10}^2} \int d^{10}x\,  \sqrt{-\det G}\, e^{-2\gF}\, 
\Big\{ 
R(\go_+)  + 4\, \der_M \gF \der^M \gF 
- \frac 12\, \big| H_3 \big|^2 
- \frac {\ga'}{4}\, \text{tr} \big| F_2 \big|^2
\Big\}~, 
} 
where $\gk_{10}^2 = \sfrac 12 (2\pi)^7 (\ga')^4$, $\gF$ is the dilaton and $G_{MN}$ the ten-dimensional metric. Its curvature scalar $R(\go_+)$ involves the spin-connection with torsion $\go_+ = \go + \sfrac 12\, H$. $F_2$ denotes the non-Abelian gauge field strength. The field strength of the Kalb-Ramond field reads 
\equ{ 
H_3 = d B_2 - \frac{\ga'}4\, X_3~, 
\qquad 
d X_3 = X_4 = \textsc{tr} F_2^2 - \textsc{tr} R_2(\go_+)^2~. 
}
Here $\textsc{tr}$ denotes the trace in the fundamental\footnote{We follow a convention where we call the lowest-dimensional irreducible representation of a Lie algebra the ``fundamental'' representation, even for groups other than SU($N$). Details concerning our trace conventions can be found in Appendix~\ref{sc:Traces}.} (vector) representation of an SO($N$) group. Since the adjoint representation of E$_8$ is its fundamental we use $\textsc{tr} F_2^2 = \sfrac 1{30} \Tr F_2^2$. In cases where we need to distinguish the two gauge factors of the E$_8\times$E$_8$ or SO(16)$\times$SO(16) we denote their field strengths etc.\ by $F_2 = (F_2^{\prime}, F_2^{\prime\prime})$, e.g.: $\textsc{tr} F_2^2 = \textsc{tr} F_2^{\prime 2} + \textsc{tr} F_2^{\prime\prime 2}$. With the trace $\Tr$ we denote the trace over all charged Majorana-Weyl fermions. If we wish to make this more explicit, we denote by $\Tr_\text{E$_8\times$E$_8$}$ and $\Tr_\text{SO(32)}$ the trace in the adjoint of E$_8\times$E$_8$ or SO(32) theories, respectively. While for the non-supersymmetric SO(16)$\times$SO(16) we have 
\equ{ \label{trSO16} 
\Tr_\text{\cancel{SUSY}} F_2^p= \big[\tr_{(\rep{128},\rep{1})}+ \tr_{(\rep{1},\rep{128})} - \tr_{(\rep{16},\rep{16})}\big](F_2^p)~, 
\qquad 
\Tr_\text{\cancel{SUSY}} F_2^2 = 0~, 
}
keeping track of the relative chiralities of the fermions. 
The latter equation follows for SO(16) from $\tr_{\rep{128}} F_2^2 = 16\, \textsc{tr} F^2$. Moreover, since SO(16)$\times$SO(16) is a subgroup of both E$_8\times$E$_8$ and SO(32), we can compare traces in both supersymmetric theories with the non-supersymmetric one, provided that we restrict $F_2$ to the adjoint of SO(16)$\times$SO(16). By considering the branching rules of the adjoints of E$_8\times$E$_8$ and SO(32) into irreducible representations of SO(16)$\times$SO(16), we find that 
\equ{ \label{trRelation}
\Tr_\text{\cancel{SUSY}} F_2^p = \Tr_\text{E$_8\times$E$_8$} F_2^p - \Tr_\text{SO(32)} F_2^p~. 
}

Anomaly cancellation requires the so-called Green-Schwarz mechanism~\cite{Green:1984sg,Green:1984bx} which involves the term 
\equ{ \label{GSaction} 
S_\text{GS} = 
\frac 1{24 (2\pi)^5\ga'}
\int 
B_2\, X_8~, 
}
in the normalization established in~\cite{Blumenhagen:2006ux}. The polynomial $X_8$ is given by (see e.g.~\cite{gsw_2}) 
\begin{subequations} 
\label{X8s} 
\equa{ \label{X8susyE8} 
X_8^\text{E$_8\times$E$_8$}   & =  
~~ \frac 1{24}\, \Tr F_2^4 - \frac 1{7200}\, (\Tr F_2^2)^2 
- \frac 1{240}\, \Tr F_2^2\, \textsc{tr} R_2^2  
+ \frac 18\, \textsc{tr} R_2^4 + \frac 1{32}\, (\textsc{tr} R_2^2)^2~, 
\\[1ex]  \label{X8susySO32} 
X_8^\text{SO(32)}  & =  
- \frac 1{24}\, \Tr F_2^4 + \frac 1{7200}\, (\Tr F_2^2)^2 
+ \frac 1{240}\, \Tr F_2^2\, \textsc{tr} R_2^2  
- \frac 18\, \textsc{tr} R_2^4 - \frac 1{32}\, (\textsc{tr} R_2^2)^2~, 
\\[1ex]  \label{X8nonsusy} 
X_8^{\text{\cancel{SUSY}}} 
& = 
~~ \frac 1{24}\, \Tr F_2^4~, 
}
\end{subequations} 
for the supersymmetric E$_8\times$E$_8$, SO(32), and non-supersymmetric SO(16)$\times$SO(16) theories, respectively. Note that in the non-supersymmetric theory the curvature two-form $R_2$ does not appear, hence in this theory the pure (irreducible and reducible) and mixed gravitational anomalies all cancel automatically. Here we have chosen the chiralities of the gravitino, dilatino and gauginos in the E$_8\times$E$_8$ and SO(32) precisely opposite to each other; this accounts for the relative sign between the $X_8$'s of both supersymmetric theories. With this convention, one obtains the relation 
\equ{ 
X_8^{\text{\cancel{SUSY}}}  = X_8^\text{E$_8\times$E$_8$} + X_8^\text{SO(32)}~ 
}
between the three eight-forms $X_8$ for the three heterotic theories. 
This result arises by making use of the identity~\eqref{trRelation} and that the quadratic trace \eqref{trSO16} vanishes.

\subsection{Heterotic lattices}

The three string theories contain both massless states and states at arbitrary high mass levels, which can be efficiently encoded in lattices. In Table~\ref{tb:Lattices} we give the full lattices on which the three heterotic theories are constructed. In particular, this displays the root lattice of the gauge group to which their sixteen-component roots $\alpha = (\alpha^I) = (\alpha^1,\ldots,\alpha^{16})$ belong. The lattices of the E$_8\times$E$_8$ and SO(32) theories show that these theories are supersymmetric at every mass level separately, while the SO(16)$\times$SO(16) is not supersymmetric at any mass level. 

In ten dimensions the overall notion of positive chirality is of course just a convention. 
We have chosen the spinorial lattices of E$_8\times$E$_8$ and SO(32) such that their chiralities are compatible with the non-supersymmetric twists to the SO(16)$\times$SO(16) theory as we discuss above. In particular, we take the gauginos $\gPs_{+} = \gPs_{({\ga_0}/2,\ldots,{\ga_3}/2)}$ in the E$_8\times$E$_8$ theory to have positive chirality, i.e.\ the product of the signs $\ga_0\cdot\ldots\cdot\ga_3=+1$, while the gauginos of the SO(32) theory are taken to be co-spinors $\gPs_{-} = \gPs_{({\ga_0}/2,\ldots,{\ga_3}/2)}$ with $\ga_0\cdot\ldots\cdot\ga_3=-1$.

\begin{table}[t!]
\centering
 \begin{tabular}{|c||c||c|c|}
 \hline
 \multicolumn{4}{|c|}{\bf Lattices in heterotic string theories}  \\
 \hline
{\bf~~~~N=1, E$\boldsymbol{_8\times}$E$\boldsymbol{_8}$~~~~} & 
{\bf~~~~N=1, SO(32)~~~~}   & 
\multicolumn{1}{|c}{\bf N=0, SO(16)$\boldsymbol{\times}$SO(16)} & $\boldsymbol{\supset}$ {\bf massless states} 
 \\\hline\hline 
 {\bf V}$_4$ $\otimes$ {\bf R}$_8$ $\otimes$ {\bf R}$_8$ &
 \multirow{2}{*}{{\bf V}$_4$ $\otimes$ {\bf R}$_{16}$} & 
 {\bf V}$_4$ $\otimes$ {\bf R}$_8$ $\otimes$ {\bf R}$_8$ &
 $(\rep{1},\rep{120}) + (\rep{120},\rep{1})$ 
 \\
 {\bf V}$_4$ $\otimes$ {\bf S}$_8$\, $\otimes$ {\bf S}$_8$\, &
 &
 {\bf V}$_4$ $\otimes$ {\bf S}$_8$\, $\otimes$ {\bf S}$_8$\,   
 & 10D gauge fields 
 \\\hline
  {\bf V}$_4$ $\otimes$ {\bf S}$_8$\, $\otimes$ {\bf R}$_8$ &
  \multirow{2}{*}{{\bf V}$_4$ $\otimes$ {\bf C}$_{16}$} &
  {\bf R}$_4$ $\otimes$ {\bf C}$_8$ $\otimes$ {\bf V}$_8$ &
  \multirow{2}{*}{--}
 \\
  {\bf V}$_4$ $\otimes$ {\bf R}$_8$ $\otimes$ {\bf S}$_8$\, &
  &   
  {\bf R}$_4$ $\otimes$ {\bf V}$_8$ $\otimes$ {\bf C}$_8$  & 
 \\\hline\hline 
 {\bf S}$_4$\, $\otimes$ {\bf R}$_8$\, $\otimes $ {\bf R}$_8$ &
 \multirow{2}{*}{{\bf C}$_4$\, $\otimes$ {\bf R}$_{16}$} &
 {\bf S}$_4$\, $\otimes$ {\bf S}$_8$\, $\otimes$ {\bf R}$_8$  & 
 $(\rep{1},\rep{128}) + (\rep{128},\rep{1})$ 
 \\ 
 {\bf S}$_4$\, $\otimes$ {\bf S}$_8$ $\otimes $ {\bf S}$_8$\, &  
 &
 {\bf S}$_4$\, $\otimes$ {\bf R}$_8$ $\otimes$ {\bf S}$_8$  &
 10D spinors
 \\\hline
  {\bf S}$_4$ $\otimes$ {\bf S}$_8$ $\otimes$ {\bf R}$_8$ &  
   \multirow{2}{*}{{\bf C}$_4$\, $\otimes$ {\bf C}$_{16}$} & 
  {\bf C}$_4$ $\otimes$ {\bf V}$_8$ $\otimes$ {\bf V}$_8$  & 
 $(\rep{16},\rep{16})$  
 \\
  {\bf S}$_4$ $\otimes$ {\bf R}$_8$\, $\otimes$ {\bf S}$_8$\, &
  & 
  {\bf C}$_4$ $\otimes$ {\bf C}$_8$ $\otimes$ {\bf C}$_8$ & 
  10D co-spinors  
 \\\hline
 \end{tabular}
\caption{\label{tb:Lattices}
The different lattices that occur in the eight (or four) different sectors of the supersymmetric E$_8\times$E$_8$, SO(32) and the non-symmetric SO(16)$\times$SO(16) heterotic string theories. {\bf V}, {\bf R}, {\bf S} and {\bf C} refer to the vector, root, spinor and co-spinor lattices, respectively. Their subscripts indicate the dimension of these lattices. Consequently, the first lattices in the tensor products classify the states as spacetime bosons ({\bf V}, {\bf R}) and fermions ({\bf S}, {\bf C}), respectively, while the remainders correspond to the various gauge representation lattices.  
We have chosen the chiralities of the spinorial lattices of E$_8\times$E$_8$ and SO(32) such that they are compatible with those of the SO(16)$\times$SO(16) theory. 
}
\end{table}

\subsection{Non-supersymmetric twists}

The three heterotic theories are closely related on the level of their respective worldsheet theories. For example, the partition function of the E$_8\times$E$_8$ and SO(16)$\times$SO(16) theories are identical up to some different choices of GSO phases. It is well-known that the  E$_8\times$E$_8$ and SO(32) theories are T-dual when compactified on a circle with appropriately chosen Wilson lines. Moreover, the SO(16)$\times$SO(16) can be obtained from either the E$_8\times$E$_8$ or SO(32) theory by supersymmetry-breaking twists: 
\begin{subequations} 
\enums{
\item[I.]
A $\Intr_2$ orbifolding of the E$_8\times$E$_8$ string with twist $v_0 = (0, 1^3)$ and gauge shift $V_0 = (1,0^7)(\sm1,0^7)$: 
\equ{ \label{E8twist}
A_M^A \ra A_M^A~, 
\qquad 
A_M^X \ra - A_M^X~, 
\quad\qquad 
\gPs_+^A \ra - \gPs_+^A~, 
\qquad 
\gPs_+^X \ra \gPs_+^X~. 
}
\item[II.] A $\Intr_2$ orbifolding of the SO(32) string with twist $v_0 = (0, 1^3)$ and gauge shift $V'_0 = (1,0^7)(\sm\sfrac 12, \sfrac 12^7)$: 
\equ{ \label{SO32twist} 
A_M^A \ra A_M^A~, 
\qquad 
A_M^Y \ra - A_M^Y~, 
\quad\qquad 
\gPs_-^A \ra - \gPs_-^A~, 
\qquad 
\gPs_-^Y \ra \gPs_-^Y~. 
}
}
\end{subequations} 
The gauge fields $A_M^A$ of the SO(16)$\times$SO(16) theory, labeled by $A$, are part of the untwisted sector in either case. The additional fermionic matter states can partially be interpreted as untwisted and partially as twisted states: In the non-supersymmetric orbifold of the E$_8\times$E$_8$ the $(\mathbf{128},\mathbf{1}) + (\mathbf{1},\mathbf{128})$ spinor states $\gPs_+^X$, labeled by $X$, are untwisted while $(\mathbf{16},\mathbf{16})$ co-spinor $\gPs_-^Y$, labeled by $Y$, are twisted. For the non-supersymmetric orbifold of the SO(32) this assignment is precisely the other way around. All these relations between the three ten-dimensional heterotic string theories are schematically indicated in Figure~\ref{fg:RelationHetStrings}.

The actions of the non-supersymmetric twists can be elegantly represented on the ten-dimensional vector multiplets using  N=1 four-dimensional superspace language~
\cite{Marcus:1983wb,ArkaniHamed:2001tb}. One decomposes the gauge fields as $A_M = (A_\gm, A_a, A_\ua)$ with four-dimensional and complexified six-dimensional indices $\gm=0,\ldots4$, $a,\ua =1,2,3$ (suppressing the gauge index for now). Their components are contained in vector superfields $\cV$ and chiral superfields $Z_a$ such that 
\equ{ 
\sfrac 12\, [\bD_\dga, D_\ga] \cV\big| = \gs^\gm_{\dga\ga}\, A_\mu~, 
\qquad 
Z_a\big| = A_a~.
}
where $|$ denotes setting all Grassmann variables $\gth$ to zero. The ten-dimensional gaugino components $\gPs_\pm$ are then represented as 
\begin{subequations} 
\equ{ 
\gPs_{(\sfrac\ga2,\sfrac\ga2^3)\phantom{\sm}}~:~ W_\ga \big| = \gl_\ga~, 
\qquad
\gPs_{(\sfrac\ga2, \undr{\sfrac\ga2,\sm\sfrac\ga2^2})}~:~\sfrac1{\sqrt 2}\, D_{\ga} Z_a \big| = \gps_{a\ga}~, 
\\[1ex]  
\gPs_{(\sfrac\ga2,\sm\sfrac\ga2^3)}~:~ W_\ga \big| = \gl_\ga~, 
\qquad
\gPs_{(\sfrac\ga2, \undr{\sm\sfrac\ga2,\sfrac\ga2^2})}~:~\sfrac1{\sqrt 2}\, D_{\ga} Z_a \big| = \gps_{a\ga}~, 
}
\end{subequations} 
for E$_8\times$E$_8$ and SO(32), respectively. 
Here $W_\ga = -\sfrac 14\, \bD^2 (e^{-V} D_\ga e^{V})$ is the superfield strength of the four-dimensional superspace and the underline denotes permutation of the entries. 
The actions \eqref{E8twist} and \eqref{SO32twist} of the non-supersymmetric twists of the E$_8\times$E$_8$ and SO(32) theories on these superfields are given by
\begin{subequations} 
\enums{
\item[I.] The $\Intr_2$ twist of the E$_8\times$E$_8$ gauge multiplets: 
\equ{ \label{E8twistSF}
(\gth^\ga, \bgth^\dga) \ra -(\gth^\ga, \bgth^\dga)~, 
\qquad 
\cV^A \ra \cV^A~, 
\quad 
\cV^X \ra - \cV^X~, 
\quad 
Z_a^A \ra Z_a^A~, 
\quad 
Z_a^X \ra -Z_a^X~. 
}
\item[II.] The $\Intr_2$ twist of the SO(32) gauge multiplets: 
\equ{ \label{SO32twistSF} 
(\gth^\ga, \bgth^\dga) \ra -(\gth^\ga, \bgth^\dga)~, 
\qquad 
\cV^A \ra \cV^A~, 
\quad 
\cV^Y \ra - \cV^Y~, 
\quad 
Z_a^A \ra Z_a^A~, 
\quad 
Z_a^Y \ra -Z_a^Y~. 
}
}
\end{subequations} 
The simultaneous reflection of all Grassmann variables ensures that the SO(16)$\times$SO(16) gaugino components are all projected out by this non-supersymmetric twist. Consequently, if we want to use four-dimensional N=1 superfields to represent the non-supersymmetric SO(16)$\times$SO(16) theory, we have the following lowest non-vanishing components,  
\begin{subequations} 
\equ{
\sfrac 12\, [\bD_\dga, D_\ga] \cV^A\big| = \gs^\gm_{\dga\ga}\, A^A_\mu~,
\quad 
Z^A_a\big| = A^A_a~, 
\\[1ex] 
W_\ga^X\big| = \gl_\ga^X~,
\quad 
D_\ga Z_a^X\big| = \gps_{\ga a}^X~, 
\quad 
W_\ga^Y\big| = \gl_\ga^Y~,
\quad 
D_\ga Z_a^Y\big| = \gps_{\ga a}^Y~, 
}
\end{subequations} 
of the superfields, $\cV^A$, $Z_a^A$, $Z^X_a$, and $Z^Y_a$ defined above. 
In addition, the SO(16)$\times$SO(16) adjoint vector and chiral multiplets, $\cV^A$ and $Z^A_a$, may contain non-vanishing auxiliary fields, $D^A$ and $F^A_a$, respectively. Just as in the supersymmetric theories, using their algebraic equations of motion these auxiliary components can be expressed in terms of the dynamical fields in the theory. 
In other words in the non-supersymmetric theory the superfields define very convenient short-hand notations.

\section{Smooth backgrounds}
\label{sc:SmoothBackgrounds} 

When one starts from a non-supersymmetric theory, there seems to be no reason to consider backgrounds that would preserve supersymmetry by themselves. However, as was pointed out in~\cite{Blaszczyk:2014qoa} it may be very convenient to consider such backgrounds as there are more computational tools available.

We focus primarily on line bundle backgrounds, which only satisfy the Bianchi identities in cohomology. This means that one is not really working on a smooth Calabi-Yau manifold, but rather on a more complicated torsion manifold. The corrections to the BPS equations, the so-called Strominger system~\cite{Strominger:1986uh,Becker:2003yv,Becker:2003sh,Becker:2006et}, give the next-to-leading corrections in the $\ga'$-expansion. Given that we only work to leading order in $\ga'$, we will ignore complications due to torsion in the following.

\subsection{Calabi-Yau manifolds}
\label{sc:CYgeom}

A very crude characterization of a Calabi-Yau manifold is given by its Hodge numbers $h_{11}$ and $h_{21}$ which count the number of independent closed (1,1)- and (2,1)-forms or their corresponding hyper-surfaces.

\subsubsection*{Topological data}

In more detail, any Calabi-Yau manifold $X$ contains a set of complex codimension one hypersurfaces called divisors. A large class of so-called toric Calabi-Yau spaces are constructed as hyper-surfaces in some toric ambient space. Toric divisors of this ambient space are defined by simply setting one of the homogeneous coordinates to zero, i.e.\ $\cD_a := \lbrace z_a = 0\rbrace$. In general these divisors are dependent, which means that there are various linear equivalence relations among them. We denote a basis of $h_{11}$ independent elements constructed out of the divisors $\cD_a$ by $\lbrace D_i\rbrace$, and a basis of $h_{11}$ curves by $\lbrace C_i \rbrace$. In this work we mostly focus on so-called ``favorable'' Calabi-Yau spaces for which this basis of divisors descends from the hyperplane classes of the projective ambient space. 

In terms of the aforementioned basis of divisors we have the triple intersection numbers and the second Chern classes evaluated on the $D_i$,
\equ{ 
\kappa_{ijk} = \int \widehat D_i \widehat D_j \widehat D_k~,
\qquad 
c_{2i} = \int_{D_i} c_2 = 
- \int_{D_i} \frac12 \tr \Big( \frac{\mathcal{R}}{2\pi} \Big)^2~. 
}
Here the curvature two-form $\mathcal{R}$ is SU(3)-valued, so that the trace is evaluated in the fundamental of this holonomy group. 
In the first expression the closed but not exact (1,1)-forms associated to the divisors $D_i$ are denoted by $\widehat D_i$; similarly we denote by $\widehat C_i$ the (2,2)-forms associated to the curves $C_i$. In fact, we may take them to be harmonic.  
Moreover, it is in principle always possible to construct an integral basis of curves and divisors, $\lbrace C_i \rbrace$ and $\lbrace D_i\rbrace$, such that 
\equ{ \label{IntegralBasis} 
\int_{C_i} \widehat D_j = \int_{D_i} \widehat C_j = \delta_{ij}~.  
}
Finally, we denote the independent (2,1) two-forms by $\widehat\go_p$ with $p=1,\ldots, h_{21}$.

\subsubsection*{Classical volumes}

The K\"ahler form $J$ can be expanded in the $\widehat D_i$ basis as 
\equ{
J = a_i\, \widehat D_i~, 
}
in terms of the $h_{11}$ K\"ahler moduli $a_i$. 
The fundamental form $J$ is used to determine the volumes of any curve $C$, divisor $D$ and the manifold $X$ itself: 
\equ{ \label{Volumes} 
\text{Vol}(C) = \int_C J~, 
\qquad 
 \text{Vol}(D) = \sfrac 12 \int_D J^2~, 
\qquad\quad 
\text{Vol}(X) = \sfrac 16 \int_X  J^3~.
\qquad 
}

In the integral basis~\eqref{IntegralBasis} we know that all the moduli satisfy $a_i > 0$ in the K\"ahler cone to guarantee that all curves $C_i$ are effective, i.e.\ have positive volume. 
In fact, we work in the large volume approximation where volumes are much larger than the string scale such that we can reliably neglect higher order $\alpha^\prime$-corrections.
The volumes of curves, divisors and $X$ read 
\equ{ \label{VolumesKahlerParameters}
\text{Vol}(C_i) = a_i~, 
\qquad 
 \text{Vol}(D_i) =  \sfrac 12\, \kappa_{ijk}\, a_j a_k~, 
\qquad 
\text{Vol}(X) = \sfrac 16\, \kappa_{ijk }\, a_i a_j a_k~.
}
Consequently, one has $\text{Vol}(X) = \sfrac 13\, \text{Vol}(C_i)\, \text{Vol}(D_i)$.

\subsubsection*{Complete intersection Calabi-Yau manifolds}

Manifolds known as Complete Intersection Calabi-Yau (CICY) manifolds are described in terms of intersecting hypersurfaces in projective ambient spaces. All smooth CICYs have been classified in \cite{Candelas:1987kf} and are available online \cite{CYweb}. Their discrete symmetries have been classified in \cite{Braun:2010vc}. Here we focus on the subclass of favorable CICYs, which means that all CICY divisors can be pulled back from the hyperplane divisors of the projective ambient spaces $\otimes_a \mathbb{P}^{k_a}$. Consequently, $a$ runs from $1$ to $h_{11}$. CICYs can be described most easily in terms of their configuration matrix  $\Gamma = (\Gamma_{aA})$. Each row, labeled by $a$, corresponds to one ambient space $\mathbb{P}^{k_a}$ factor and each column, labeled by $A$, corresponds to one polynomial that defines a hypersurface in the ambient space. Thus, an entry $\Gamma_{aA}$ specifies the scaling of the $A^\text{th}$ polynomial under the projective scale factor of the $a^\text{th}$ projective ambient space factor. Since each polynomial imposes one constraint, we find that $A$ runs from $1$ to $\sum_a k_a-3$ for a CY 3-fold. The first Chern class of a CICY vanishes if
\equ{ 
\sum_A \Gamma_{aA} = k_a +1\,.
}
In this way, the ambient space follows uniquely from the configuration matrix.

For the calculation of the intersection numbers $\kappa_{ijk}$ and the total Chern class of CICY manifolds we used the methods introduced in \cite{Hosono:1994ax}:
\equ{ 
 \kappa_{ijk}=
 \prod_{e} \frac{1}{k_e!}\, \frac{\partial^{k_e}}{\partial D_e^{k_e}}
 \left[
 \prod_{A} \Big( \sum_{a}\Gamma_{aA} D_a\Big) \, c(X)\,   
   D_i D_j D_k 
 \right]_{D=0}\,,~ 
 c(X) =  \frac{\dsp \prod_{a} \Big(1+D_a\Big)^{k_a+1}}
 {\dsp \prod_{B}\Big(1+\sum_{b}\Gamma_{bB}D_b\Big)}\,, 
}
where $A,B = 1,\ldots, \sum_a k_a-3$ and $a,b,\ldots =1,\ldots, h_{11}$.

\subsubsection*{Free quotients}

Next, we consider a discrete, freely acting symmetry group $\gG$ of finite order $|\gG|$ that acts on the coordinates of $X$ as $z \rightarrow g z$. 
In this work we will assume that $\gG$ consists of a single $\Intr_N$ factor only. Since $\gG$ is assumed to act freely, the quotient
\equ{ 
\widetilde{X} = X / \gG
}
is again a smooth, but not simply-connected, Calabi-Yau manifold. 

The action of the discrete group $\gG$ can be described in terms of the action on the ambient space coordinates. In order to be able to mod out such an action, one has of course to ensure that the Calabi-Yau geometry admits such a symmetry. For example, this poses constraints on the ambient space and the polynomials whose intersections define the CICY.  This typically means that some complex structure deformations, counted by $h_{21}$, are frozen. There are essential three ways in which $\gG$ can act  \cite{Braun:2010vc}:
\begin{enumerate}
\item homogeneous coordinates obtain phases,
\item homogeneous coordinates within each $\mathbb{P}^{N}$ factor are permuted, 
\item or complete $\mathbb{P}^{N}$ factors are permuted  among each other.
\end{enumerate}
In the first case the ambient space divisors  $\cD_a = \lbrace z_a=0 \rbrace$ are invariant. The second type of action permutes them among each other, but the corresponding divisor class remains invariant. Hence, if one had chosen it as one of the divisor basis elements $D_i$, then it remains inert. In contrast, in the third case one has to form invariant linear combinations of divisors $\lbrace D_i\rbrace$ of $X$. This means that in the third case $h_{11}$ is reduced as well.

\subsection{Line bundles on Calabi-Yaus}
\label{sc:GaugeBundle}

For the gauge background $\mathcal{F}$ we make the simple ansatz that the vector bundle $\mathcal{V}$ is given by a sum of line bundles (see e.g.~\cite{Nibbelink:2007rd,Nibbelink:2009sp}) 
\equ{ \label{LineBundleFlux} 
 \frac{\mathcal{F}}{2\pi} = \widehat D_i\, H_i~, 
 \qquad 
 H_i = V_i^I\, H_I~, 
 }
which are embedded in the Cartan subalgebra of the ten-dimensional gauge group $\mathcal{G}$. This gauge background is characterized by a set of bundle vectors $V_i =(V_i^I)$, one for each divisor (1,1)-form $\widehat D_i$. The Cartan generators, $H_I$, $I=1,\ldots,16$, are assumed to be normalized such that\footnote{The trace itself is normalized as the trace over the fundamental representation of SU-groups.}
\equ{ \label{CartanNorm} 
\tr (H_I H_J) = \delta_{IJ}~. 
}
Consequently, $\tr( H_i H_j) = V_i\cdot V_j$, where $\cdot$ is the standard (euclidean) inner product of two vectors with sixteen components. The unbroken subgroup $\mathcal{H}$ of the ten-dimensional gauge group $\mathcal{G}$ is generated by this Cartan subalgebra augmented with the creation and annihilation operators associated with the roots $\alpha$ that are perpendicular to all line bundle vectors, $V_i \cdot \alpha = 0$, for all $i=1,\ldots,h_{11}$. For the E$_8\times$E$_8$ and SO$(16)\times$SO$(16)$ theories we decompose these bundle vectors w.r.t. observable and hidden gauge group factors as $V_i = (V_i^{\prime}, V_i^{\prime\prime})$ and similarly for other quantities where applicable.

Any line bundle background is subject to a number of consistency conditions, which we list in the following:

\subsubsection*{Flux quantization}

The bundle vectors are subject to flux quantization conditions, which ensure that  
\begin{equation}
\int_{C} \frac{\mathcal{F}}{2\pi}~, 
\end{equation}
evaluated on any state $| p \rangle$ in the full string spectrum, is integral for all curves $C$. 
 
If $\{D_k\}$ is an integral basis of divisors satisfying \eqref{IntegralBasis},  this amounts to requiring that all line bundle vectors lie on the lattices 
\equ{ 
 \gL_\text{E$_8\times$E$_8$} =  (\mathbf{R}_8 \oplus \mathbf{S}_8) \otimes (\mathbf{R}_8\oplus \mathbf{S}_8)~, 
\qquad 
\gL_\text{SO(32)} = \mathbf{R}_{16} \oplus \mathbf{S}_{16}~, 
}
in the cases of the E$_8\times$E$_8$ or SO(32) heterotic string, respectively.  
The flux quantization in the non-supersymmetric SO(16)$\times$SO(16) theory requires the bundle vectors to lie on the lattice  
\begin{equation}\label{GaugVectors_on_lattice}
 \gL_\text{\cancel{SUSY}} =  \big(\mathbf{R}_8 \otimes \mathbf{R}_8) \oplus \big(\mathbf{S}_8 \otimes \mathbf{S}_8)~, 
\end{equation}
which is contained in both  E$_8\times$E$_8$ and SO(32) lattices. 
Consequently, any allowed set of bundle vectors of the SO(16)$\times$SO(16) theory also represents an admissible set for either of the supersymmetric heterotic theories.

\subsubsection*{Bianchi identities}

The Bianchi identities for the $B$-field constitute further consistency conditions on the line bundle gauge background: 
\equ{ \label{BIs} 
\tr \Big( \frac{\mathcal{F}}{2\pi} \Big)^2 - 
\tr \Big( \frac{\mathcal{R}}{2\pi} \Big)^2  
= 
N_i\, \widehat C_i
}
in cohomology, i.e.\ when integrated over any divisor $D$ of $X$. In the integral basis~\eqref{IntegralBasis} we can interpret $N_i$ as the five-brane charge associated to the five-brane wrapping the curve ${C}_i$: 
\begin{subequations}
\equ{
N_i = N^{\prime}_i + N^{\prime\prime}_i~,
\qquad 
N^{\prime}_i =  \kappa_{ijk}\, V^{\prime}_j \cdot V^{\prime}_k + c_{2i}~,
\quad 
N^{\prime\prime}_i =  \kappa_{ijk}\, V^{\prime\prime}_j \cdot V^{\prime\prime}_k + c_{2i}~,\label{eq:BraneChargeSplit}
}
for the  E$_8\times$E$_8$ or SO(16)$\times$SO(16) theories and 
\equ{ 
N_i =  \kappa_{ijk}\, V_j \cdot V_k + 2\, c_{2i}~,\label{eq:BraneChargeFull}
} 
\end{subequations}
for the SO(32) theory, respectively. 
When all $N_i \geq 0$ the configuration of five-branes preserves the same four-dimensional supersymmetry as the perturbative sector of the E$_8\times$E$_8$ or SO(32) theory. For the E$_8\times$E$_8$ theory, the non-perturbative NS5-brane spectrum involves only a number of tensor multiplets, and hence does not modify the spectrum charged under the perturbative unbroken gauge group $\mathcal{H}$. 
In contrast, for the SO(32) theory there are additional matter multiplets in bi-fundamental representations of the unbroken subgroup $\mathcal{H}$ and the non-perturbative groups Sp(2$\tN$). To the best of our knowledge, it is unknown which additional non-perturbative charged states need to be added to the SO(16)$\times$SO(16) theory; we will present a suggestion for this in Section~\ref{sc:FiveBranes}.

\subsubsection*{Donaldson-Uhlenbeck-Yau equations}

An additional requirement on the line bundle is that the DUY equations, 
\equ{\label{DUY2}  
\sfrac 12 \int J^2 \, \frac{\mathcal{F}}{2\pi} = 
 \text{Vol}(D_i)\, V_i^I = 0~, 
}
can be satisfied. At first this seems to impose a condition on the moduli only, encoded in the volumes of the divisors $D_i$, but in fact it leads to stringent restrictions on the possible line bundle vectors. This comes about because one has to ensure that the zero-vector can be obtained from a linear combination of the $V^I_i$ with positive coefficients only.

We have not included the one-loop Blumenhagen-Honecker-Weigand correction~\cite{Blumenhagen:2005ga} for the following reasons: First of all it is often possible to absorb this one loop correction by appropriately shifting the volumes of the divisors. Only when one has an embedding in both E$_8$ gauge group factors, it is not generically possible to do so. Second, the form of this correction is not known for the compactification of the non-supersymmetric SO(16)$\times$SO(16) theory.
More importantly, we expect other effects to be generated at one loop order (e.g. appearance of tachyons) in the non-supersymmetric theory. Hence, for that reason also, our analysis will focus on the weak coupling limit of the theory, where such corrections can be neglected. Note that the DUY equations are homogeneous at tree level, such that at this order one can always go to a large volume point in moduli space.

\subsubsection*{Equivariant line bundles} 

By accompanying the action of the freely acting symmetry $\gG$, which was used to obtain the non-simply-connected Calabi-Yau $\widetilde{X} = X/\gG$ from the Calabi-Yau $X$, with an action on the gauge degrees of freedom, 
\equ{ 
A(z) \rightarrow A(g\, z) = W_g\, A(z)\, W_g^{-1}~, 
} 
we can induce a further gauge symmetry breaking: $\mathcal{H} \rightarrow \widetilde{\mathcal{H}}$. In practice, such a freely acting Wilson line $W_g$ induces a non-local gauge symmetry breaking, typically chosen such that a GUT subgroup of $\mathcal{H}$ in the upstairs description is broken down to the SM group in the downstairs picture. However, before we divide out a freely acting symmetry in a heterotic theory, we need to make sure that the bundle is equivariant under (i.e.\ compatible with) the action: 
\equ{ 
\mathcal{F}(z) \rightarrow \mathcal{F}(g\, z) = 
W_g\, \mathcal{F}(z)\, W_g^{-1}~, 
}
for all $g \in \gG$. We assume that the generator of the freely acting symmetry can be diagonalized simultaneously with the line bundle flux, therefore it may be written as $W_g = \exp{(2\pi i\, W^I H_I)}$. 

In the part on free quotients in Subsection~\ref{sc:CYgeom} we considered three possible $\gG$ actions on the geometry. As each of the basis divisors $D_k$ is invariant under the first two actions listed there, the line bundles constructed in the upstairs picture are automatically invariant. For the third type of freely acting symmetry, which permutes various $\mathbbm{P}^N$ factors, a simple way to ensure equivariance is to require that the corresponding gauge vectors are identical. As this reduces the number of independent line bundle vectors, this tightens the constraints on having large volume solutions to the DUY equations and satisfying the Bianchi identities without NS5-branes.

\section{Spectra of smooth Calabi-Yau compactifications} 
\label{sc:Spectrum} 

In this section we discuss methods to compute the massless spectrum of smooth compactifications of the two supersymmetric heterotic string theories and the non-supersymmetric SO(16)$\times$SO(16) theory.

\subsection{Massless charged chiral 4D spectrum}

For many applications it is sufficient to compute only the charged chiral spectrum. To this end it is convenient to use the multiplicity operator $\mathcal{N}(X)$, see e.g.\ \cite{Nibbelink:2007rd,Nibbelink:2007pn,Nibbelink:2008tv,Blaszczyk:2010db}, that counts the number of chiral states. It was obtained in~\cite{Nibbelink:2007rd} by integrating the ten-dimensional anomaly polynomial $I_{12|\rep{R}}$ over the internal Calabi-Yau manifold: 
\equ{ \label{4Danomaly} 
I_{6|\rep{R}}(X) = \int_X I_{12|\rep{R}} = \frac1{2(2\pi)^2}\, 
\tr_\rep{R}\Big[ \mathcal{N}(X)\, 
\Big( 
\frac 16\, F_2^3 - \frac 1{48}\, (\textsc{tr} R_2^2)\, F_2 
\Big)\Big]~.  
}
Here $\rep{R}$ is the representation which the ten-dimensional states are transforming in,  $R_2$ is the four-dimensional curvature two-form and $F_2$ is the gauge field strength of the unbroken gauge group $\mathcal{H}$ in four dimensions. Consequently, the multiplicity operator, 
\equ{ \label{MultiOp4D}
 \mathcal{N}(X)  = \frac1{(2\pi)^3} 
  \int_X \Big\{ \frac16\, \mathcal{F}_2^3 
  - \frac1{24}\,  \tr(\mathcal{R}_2^2)\,\mathcal{F}_2   
  \Big\}
  = \frac16\, \kappa_{ijk}\, H_i H_j H_k + \frac1{12} c_{2i}\,  H_i~, 
}
can be evaluated on each of the weights $p$ of the appropriate representations given in Table~\ref{tb:10DhetSpectra} using that $H_i(p) = V_i \cdot p$. 

The multiplicity operator was obtained in the context of the supersymmetric heterotic E$_8\times$E$_8$ string (and of course applies to the SO(32) case in a straightforward way). In Ref.~\cite{Blaszczyk:2014qoa} it was argued that this formula can also be employed for the non-supersymmetric SO(16)$\times$SO(16) theory: To determine both the chiral fermionic and bosonic spectra one has to suitably choose the representations $\rep{R}$ and keep track of the ten-dimensional chirality. To compute the number of complex scalars we take for $\rep{R}$ the adjoint representation of SO(16)$\times$SO(16), while for chiral fermions in four dimensions we take states in the spinor representation of either of the two SO(16)s, or states in the bi-fundamental $(\mathbf{16},\mathbf{16})$. Because of the opposite ten-dimensional chirality, the latter states transform in charge conjugate representations as compared to the states resulting from the spinor representations. 

The multiplicity operator can also be evaluated on the free quotient $\widetilde{X} = X/\gG$:
The downstairs multiplicity of any chiral state in the upstairs spectrum is simply given by
\equ{
 \mathcal{N}(\widetilde{X}) =  \mathcal{N}(X/\gG)
 =  \frac 1{|\gG|}\, \mathcal{N}(X)~. 
}
This can be seen as follows:  By writing $X = \bigcup_g \widetilde{X}^g$, where $\widetilde{X}^g$ denotes the image of $\widetilde{X} \supset X$ under $g\in \gG$, we can decompose the integral in~\eqref{4Danomaly} in $|\gG|$ pieces. Using the fact that $\gG$ act freely on the geometry and equivariantly on the bundle, we see that this gives $|\gG|$ equal contributions.

\subsection{Dirac and Hirzebruch-Riemann-Roch indices}

To give additional motivation for using this formula to compute the spectra for both fermions and bosons, we resort to the following index theorems: 
\items{ 
\item 
The net spectrum of chiral fermions is determined by the Dirac index~\cite{Nakahara:1990th}
\equ{\label{DiracIndex}
\text{ind}_\text{Dirac}(X,\mathcal{V})=\int_X \text{ch}(\mathcal{V})~\widehat{\text{A}}(X)~,
}
where $\text{ch}$ is the Chern character of the bundle and $\widehat{A}$ the roof-genus. 
\item 
Similarly, the net spectrum of complex bosons is characterized by the Hirzebruch-Riemann-Roch index theorem~\cite{Nakahara:1990th,Honecker:2006dt,Anderson:2012yf}
\equ{\label{HRRIndex}
\text{ind}_\text{HRR}(X,\mathcal{V})=\int_X \text{ch}(\mathcal{V})~\text{Td}(X)~,
}
which involves the Todd class $\text{Td}$ instead of the $\widehat{\text{A}}$ class. 
}
Using the splitting principle, a vector bundle $\mathcal{V}$ can be represented as $\mathcal{V}=\oplus_j \mathcal{L}_j$ where $\mathcal{L}_j$ are line bundles. Since these are one-dimensional, they are characterized completely in terms of their first Chern class. Letting $x_j=c_1(\mathcal{L}_j)$, we can express the Chern class, the Chern character, the Todd class, and the $\widehat{\text{A}}$ class as the products
\equ{ 
 \text{c}(\mathcal{V}) = \prod_j(1+x_j)~,
 \quad
 \text{ch}(\mathcal{V}) = \sum_{j}e^{x_j}~,
 \quad 
 \text{Td}(\mathcal{V})  = \prod_j\frac{x_j}{1-e^{-x_j}}~,
 \quad	 
 \widehat{A}(\mathcal{V}) = \prod_j \frac{x_j/2}{\text{sinh}(x_j/2)}~, 
}
respectively. Expanding the Todd and $\widehat{\text{A}}$ classes to third order in terms of the Chern classes, 
\begin{subequations} 
\equa{
 \text{Td}(X)   
 &= 1+\frac 12\, \text{c}_1(X) + \frac{1}{12}\Big( \text{c}_1^2(X) + \text{c}_2(X)\Big) + 
 \frac 1{24}\, \text{c}_1(X)\, \text{c}_2(X)~,  
 \\[1ex] 
\widehat{A}(X)
&=1 - \frac 1{24}\, \text{c}_1^2(X) + \frac{1}{12}\, \text{c}_2(X)~, 
}
\end{subequations} 
with $\text{c}_1(X)=\sum_j x_j$, $\text{c}_2(X)=\sum_{i>j} x_i x_j$, shows that the indices \eqref{DiracIndex} and \eqref{HRRIndex} agree when the compactification manifold has vanishing first Chern class $\text{c}_1(X)=0$. Furthermore, these indices reproduce the multiplicities determined by the multiplicity operator \eqref{MultiOp4D} evaluated on the appropriate weights.

\subsection{Beyond the chiral spectrum}
\label{sc:BeyondChiral} 

While the multiplicity operator \eqref{MultiOp4D} gives us the net chiral multiplicity of the charged states, determining the number of truly vector-like pairs is more difficult. In general one would have to compute individual dimensions of cohomology groups of appropriate wedge products of the line bundles, rather than just their alternating sum which appears in the indices. Determining the full spectrum provides a strong cross-check on the chiral spectrum determined by the multiplicity operator. Moreover, knowing the full spectrum is important in order to be able to check whether there are exotics in the spectrum that are vector-like with respect to all line bundle charges. For example, if we want to investigate whether we have the exact fermionic spectrum of the MSSM without exotics, we need to show that we have exactly $3\, |\gG|$ $\rep{10}$-plets and no $\brep{10}$-plets on the SU(5) GUT level, where $|\gG|$ is the order of the discrete Wilson line. 

In order to compute the full spectrum via cohomology group dimensions we make use of the Mathematica package \texttt{cohomcalg}~\cite{Blumenhagen:2010pv,cohomCalg:Implementation}. The idea behind this code is the following: 
The spectrum of a line bundle background on a CICY can be determined by computing the ambient space vector bundle cohomology and subsequently restricting it to the Calabi-Yau via the so-called Koszul sequence.  This is an exact sequence for a hypersurface in codimension $r$ twisted by the bundle $\mathcal{V}$
\begin{align}
\label{eq:KoszulSequence}
0\rightarrow\mathcal{V}\otimes\bigwedge^r N^*\rightarrow\mathcal{V}\otimes\bigwedge^{r-1} N^*\rightarrow\ldots\rightarrow\mathcal{V}\otimes\bigwedge^1 N^*\rightarrow \mathcal{V}\rightarrow\mathcal{V}|_{X}\rightarrow 0\,,
\end{align}
where $N^*$ is the dual normal bundle of the CICY, i.e.\ of the intersection locus of the hypersurface equations. We are interested in the last part $\mathcal{V}|_X$. By introducing auxiliary sheaves we can break this exact sequence into several short exact sequences. These give rise to long exact sequences in cohomology. We can compute the dimension of these cohomologies in the ambient space. Due to the exactness of \eqref{eq:KoszulSequence}, this allows us to determine the cohomology of the $\mathcal{V}|_X$ part we are interested in but which we cannot compute directly. Exactness implies that the alternating sum of the dimension of the cohomology groups add up to zero. Thus in cases where ``enough'' ambient space cohomology groups are trivial, i.e.\ when no more than three consecutive positions in the Koszul are non-zero, the dimensions of the cohomology groups follow uniquely. In cases where more consecutive positions are non-vanishing, all we know is that their alternating sum equals zero. Due to this, the \texttt{cohomcalg}-package does not determine the dimensions of the cohomology groups uniquely in such cases.  In order to resolve the ambiguity one has to construct explicitly the maps between the cohomology groups and work out their kernels and images.

\subsubsection*{Higgs doublet pairs in supersymmetric 4D effective theories}

The determination of the full spectrum is in particular relevant to determine the Higgs sector. Let us first consider Calabi-Yau compactifications of one of the supersymmetric heterotic strings. In addition to having at least the $3\, |\gG|$ $\brep{5}$-plets, which contain the left-handed SM quarks and leptons, we need at least one pair of $\rep{5}-\brep{5}$-plets which contain a SM Higgs candidate. (As is well-known in the MSSM one needs a pair of Higgs doublets in order to cancel anomalies induced by the Higgsinos.) 
Note that such Higgs candidate pairs of $\rep{5}-\crep{5}$-plets need to behave very differently under the freely acting symmetry $\gG$ than the $\brep{5}$ that contain the left-handed quarks and leptons: From the latter we want to retain the full $\brep{5}$-plets since their triplets correspond to the down-type quarks; merely their multiplicity should be reduce by $|\gG|$.  In contrast, $\rep{5}-\crep{5}$-plet pairs have to become split multiplets, such that the Higgs doublets survive while the triplets are projected out by the Wilson line. This means that for the remaining vector-like Higgs pair the overall multiplicity stays zero.

\subsubsection*{Higgs doublet(s) in non-supersymmetric 4D effective theories}

In non-supersymmetric four-dimensional theories it is no problem to have just a single Higgs, since it is a scalar and thus does not produce any anomalies. However, from our previous discussion we infer that it is impossible to obtain just a single Higgs doublet: 
As all chiral representations have a multiplicity which is divisible by $|\gG|$, we need to start with the lowest possible number, i.e.\ $|\gG|$, of additional $\rep{5}$-plets that can host the SM Higgs, in order to keep exactly one Higgs doublet in the downstairs spectrum. However, after dividing out the freely acting Wilson line we will then obtain one doublet and one triplet as the surviving $\rep{5}$-plet is merely branched. Alternatively, as in the supersymmetric case, we could start with a vector-like pair (such that the combined multiplicity is zero) and then divide by the freely acting symmetry such that the triplets are projected out. But then we have a pair of Higgs doublets rather than a single one. Consequently, we either have only one Higgs together with its color triplet partner or we have at least one pair of vector-like Higgs doublets.

Note that the situation is different for orbifold compactifications of the SO(16)$\times$SO(16) theory: Indeed, in~\cite{Blaszczyk:2014qoa} various orbifold models with a single Higgs doublet were obtained. This statement is not in conflict with our previous observations on smooth manifolds: The orbifold gauge shift and discrete Wilson lines in these orbifold models were constructed such that they break the gauge group directly to the SM group. In other words no freely acting symmetry was needed to break an intermediate GUT group down to the SM, but it is precisely such a freely acting symmetry that lead us to the conclusion above. When working on CY manifolds, a direct breaking of SU(5) is in general not feasible since the GUT group rank is reduced via the breaking.

\newpage

\section{Effective theories in four dimensions}
\label{sc:4DEFT}

For the compactifications of the supersymmetric E$_8\times$E$_8$ and SO(32) theories we can use the familiar N=1 superspace formalism involving the K\"ahler potential, superpotential and the gauge kinetic functions, to fully characterize the resulting effective theories in four dimensions. Moreover, as long as we neglect $\ga'$ and $g_s$ corrections we may even use this language  to efficiently describe the effective theory of Calabi-Yau compactifications of the non-supersymmetric SO(16)$\times$SO(16) string as well. However, in this case we use these functions as convenient short hands to describe specific bosonic and fermionic terms of the action. When we go beyond the leading order, this formalism breaks down since loops involving bosons and fermions are not identical anymore. However, as long as we consider Green-Schwarz interactions, that are directly related to anomaly cancellation, we can still trustworthily compute the corresponding axion couplings as in supersymmetric theories.

\subsection{Effective four-dimensional N=1 actions for \texorpdfstring{E$_8\times$E$_8$}{E8 x E8} and SO(32) compactifications}
\label{sc:EffectiveAction} 

Below we give the K\"ahler potential $\cK$, superpotential $\cW_B$ and gauge kinetic function $f$ which characterize the compactification of the supersymmetric heterotic string theories. These functions can be (partially) inferred from dimensional reductions of various terms in the ten-dimensional action~\eqref{10Daction}. We expand the Kalb-Ramond two-form and the gauge fields as 
\equ{ \label{Expansions} 
\gF = \gvf_0 + \gvf~, 
\qquad 
B_2 = b_2 + \ell_s^2\, \gb_i \, \widehat D_i~, 
\qquad 
A_1 = \cA_1 +  a_1 
} 
where $\ell_s^2 = (2\pi)^2 \ga'$ denotes the string length. The four-dimensional dilaton is denoted by $\gvf$ and the constant background value of the ten-dimensional dilaton by $\gvf_0$.  In addition, $\cA_1$ defines the gauge background with field strength $\cF_2$, and $a_1$ the four-dimensional gauge field one-forms with field strengths $F_2$. 
The Kalb-Ramond two-form can be expanded in terms of harmonic $(1,1)$-forms dual to the divisors $D_i$. The fields $\gb_i$, appearing in this expansion, are called model-dependent axions; the model-independent axion $\gb_0$ is dual to the four-dimensional two-form $b_2$: 
\equ{ \label{DualityWeylScaling}
e^{-4(\gvf+\gg_0)}*_4 d b_2 = d \gb_0~, 
\qquad 
G_{\gm\gn} = e^{2(\gvf+\gvf_0)}\,  g_{\gm\gn}~, 
}
using the four-dimensional Einstein metric $g_{\gm\gn}$, obtained from a four-dimensional Weyl rescaling, to define the Hodge dual $*_4$.  The four-dimensional Planck scale can be read off from the factor in front of the  Einstein-Hilbert term to be
\equ{  
\frac{M_P^2}{8\gp} =  
\frac1{\gk_4^{2}} = \frac1{\gk_{10}^{2}}\,  
\int \sfrac 16 J^3~, 
\qquad 
e^{4\gg_0} = e^{2\gvf_0} = \ell_s^{-6}\, 
\int \sfrac 16 J^3~. 
} 
The constants $\gg_0, \gvf_0$ are fixed such that the kinetic terms of gauge fields and their couplings to the model-independent axion $\gb_0$, obtained from the Green-Schwarz term \eqref{GSaction} involving $b_2$, can be written as 
\equ{ 
S_\text{YM} = 
\frac 14 \int d^4d^2\gth\, \textsc{tr} \Big[ S\, W^2 \Big] + \text{c.c.}~,  
}
provided that the defining components of the chiral and vector superfields contain  
\equ{ \label{S_superfields} 
S\big| \supset \frac 1{2\pi}\,  \Big[e^{-2\gvf} + i\, \gb_0\Big]~
\qquad 
\sfrac 12\, \big[\,\overline{D}_\dga, D_\ga \big] \cV\big| \supset
\gs^\gm_{\dga\ga}\, a_\gm~. 
}

This is compatible with the kinetic terms of the dilaton $\gvf$ and the axion $\gb_0$ that arise from the K\"ahler potential given by 
\equ{ \label{Kahler} 
\cK = 
- \gk_4^{-2} \, \ln \Big[ S + \overline S -  \frac 1{(2\pi)^2}\, Q^I \cV^I \Big]
- \gk_4^{-2} \, \ln \int_X \sfrac 16\, \cJ^3 
- \gk_4^{-2} \, \ln \int_X \bgO\, \gO 
- \cK_{\um m}\, \bZ_\um e^{2\, q\cdot \cV} Z_m~. 
}
The introduction of the vector multiplets $\cV^I$ in the first term is fixed by determining the gauge connection for the model-independent axion that arise from the Green-Schwarz term~\eqref{GSaction} after using the dualization~\eqref{DualityWeylScaling}. The charges, $Q_I$, in 
\equ{ \label{defQcharge}
Q^I\, F_2^I = \frac 1{24}\, \frac 1{(2\pi)^3} \int X_{2,6}~, 
}
depend strongly on the theory under consideration and are evaluated below. 
In addition, we have introduced the notation,  
\equ{  \label{KahlerThreeForm} 
\cJ = - \sfrac 12\, \Big( T_i + \overline{T}_i - \frac 1{(2\pi)^2}\, Q_i^I\, \cV^I
\Big)\, \widehat D_i~, 
\qquad 
\gO = U_p\, \widehat \go_p~,
} 
such that $\cJ| = J/(2\pi\, \ell_s^2)$ gives the K\"ahler form $J$. The (2,1)-forms were defined below \eqref{IntegralBasis} . 

Here we have defined the chiral superfields 
\equ{  \label{T_superfields} 
T_i\big| \supset  \frac 1{2\pi} \Big[ - \frac{a_i}{\ell_s^2} + i\, \gb_i \Big]~, 
\qquad 
U_p| \supset \frac 1{2\pi}\, u_p~, 
}
that involve the K\"ahler and complex structure moduli, respectively. 
The second term in~\eqref{Kahler} is determined by the kinetic terms of the model-dependent axions $\gb_i$ using~\cite{Strominger:1985ks}
\equ{ \label{InnerProductDivs} 
\int_X  \widehat D_i \wedge * \widehat D_j = 
\frac 1{4 \text{Vol}(X)}\, 
\left(\int_X J^2\, \widehat D_i \right)  \left(\int_X J^2\,  \widehat D_j\right)
- \int_X J \,\widehat D_i \widehat D_j~, 
}
where $\text{Vol}(X)$ is given in~\eqref{Volumes}. The dependence on the four-dimensional dilaton $\gvf$ dropped out via the Weyl rescaling~\eqref{DualityWeylScaling}. 

The coupling to the vector multiplets $\cV^I$ is determined by collecting the terms proportional to  $\widehat{D}_i$ in the expansion of $H_3$ using  the line bundle gauge flux~\eqref{LineBundleFlux} and~\eqref{Expansions}. Using
\equ{
H_3 \supset 2\pi\, \ell_s^2 \, \widehat D_i \, 
\Big[ d \frac{\gb_i}{2\pi} - \frac 1{(2\pi)^2}\, V_i^I\, a_1^I \Big]~,
}
we find
\equ{ \label{defQiI} 
Q_i^I = V_i^I
}
for the charge of $\gb_i/2\pi$, i.e.\ for the imaginary part of $T_i$.
The final term in~\eqref{Kahler} involves the massless chiral superfields,
\( 
Z_m| = z_m 
\), 
with charge matrix $q$. Even in Calabi-Yau compactifications of supersymmetric string theories the detailed K\"ahler potential, encoded in $\cK_{\um m}$ in~\eqref{Kahler}, is difficult to determine unless one is on special backgrounds such as orbifolds or uses the standard embedding.  

By reducing the kinetic terms of the ten-dimensional gauge fields and the cross-terms in the kinetic terms of the Kalb-Ramond field, one can extract the moduli-dependent part of the gauge kinetic function
\equ{ 
S_\text{YM} 
\supset \frac 14 \int 
 \text{Im}(T_i) \, 
\Big\{ 
\textsc{tr} \Big[\gD\! f_i^\prime\, F^{\prime\, 2}_2 + \gD\! f_i^{\prime\prime}\, F^{\prime\prime\, 2}_2 \Big]
+ \gD\! f_i^{IJ}\, F^{I}_2 F^J_2
\Big\}~. 
}
The gauge kinetic function coefficients $\gD\! f_i$ are determined by a reduction of the Green-Schwarz term \eqref{GSaction},
\equ{ \label{4DGSterms}
S_\text{GS} \supset 
\frac {1}{24(2\gp)^3}\, 
\int \gb_i\, D_i\, X_{4,4}
= \frac 1{4}\int \frac{\gb_i}{2\pi}  
\Big\{ 
\textsc{tr} \Big[\gD\! f_i^\prime\, F^{\prime\, 2}_2 + \gD\! f_i^{\prime\prime}\, F^{\prime\prime\, 2}_2 \Big]
+ \gD\! f_i^{JK}\, F^{J}_2 F^K_2
+ \gD_i\, \textsc{tr} R^2 
\Big\}
~. 
}
These charges and coefficients determine the factorization of the anomaly polynomial in four dimensions, 
\equ{ 
4\, I_6 =  Q^I \, F_2^I\, \Big\{ \textsc{tr} F_2^2 - \textsc{tr} R_2^2 \Big\} + 
Q_i^I F_2^I\, \Big\{ 
\textsc{tr} \Big[\gD\! f_i^\prime\, F^{\prime\, 2}_2 + \gD\! f_i^{\prime\prime}\, F^{\prime\prime\, 2}_2 \Big]
+ \gD\! f_i^{JK}\, F^{J}_2 F^{K}_2
+ \gD_i \, \textsc{tr} R_2^2 
\Big\}~.  
}
Here $I_6$ is the four-dimensional anomaly polynomial computed directly using the multiplicity operator. Given the prefactor $1/4$ in \eqref{4DGSterms}, we have a normalization factor $4$ in the anomaly factorization formula. 

The explicit expressions for the charges, $Q^I_i$, $Q^I$, and the coefficients, $\gd_\text{GS}$, $\gD\! f_i$, are theory-dependent: 
\subsubsection*{Supersymmetric E$_8\times$E$_8$ theory:} 
\begin{subequations} 
\equa{ \label{qE8} 
Q_i^{I^\prime} &= V^{\prime I}_i~, 
\qquad\qquad\qquad\qquad~ 
Q_i^{I^{\prime\prime}} = V^{{\prime\prime}I}_i~, 
\\[1ex]
Q^{I^{\prime}} &= 
\sfrac 1{6}\, V_i^{{\prime}I}\, \Big(N_i^\prime - \sfrac 1{2}\, N_i^{\prime\prime} \Big)~, 
\quad~~ 
Q^{I^{\prime\prime}} = 
\sfrac 1{6}\, V_i^{{\prime\prime}I}\, \Big(N_i^{\prime\prime} - \sfrac 1{2}\, N_i^{\prime} \Big)~, 
\\[1ex] 
\gD\! f_i^{\prime} &=  \sfrac 1{6}\,  
\Big( N_i^\prime - \sfrac 1{2}\, N_i^{\prime\prime} \Big)~,  
\qquad~~
~\,\gD\! f_i^{\prime\prime} = \sfrac 1{6}\, 
\Big(N_i^{\prime\prime} - \sfrac 1{2}\, N_i^{\prime}  \Big)~, 
\\[1ex]
\gD\! f_i^{J^\prime K^\prime} & = 
\sfrac 23\,  \gk_{ijk}\, V_{j}^{\prime J} \,V_k^{\prime K}~, 
\qquad 
\gD\! f_i^{J^{\prime\prime} K^{\prime\prime}}  =  
\sfrac 23\, \gk_{ijk}\, V_j^{\prime\prime J} \, V_k^{\prime\prime K}~,  
\qquad 
\gD\! f_i^{J^\prime K^{\prime\prime}} = 
-\sfrac 13\, \gk_{ijk}\, V_j^{\prime J} V_k^{\prime\prime K}~, 
\\[1ex] 
\gD_i &= - \sfrac 1{24}\, \big( \gk_{ijk}\, V_j \cdot V_k + c_{2i}\big)~. 
}
\end{subequations} 
\subsubsection*{Supersymmetric SO(32) theory:} 
\begin{subequations}
\equa{ \label{qSO32}
- Q^I_i &= V^I_i~,
\qquad 
-Q^I = \sfrac 16\, \gk_{ijk}\, \textsc{tr}(H_i H_j H_k H_I) + \sfrac 1{12}\, c_{2i}\, V_i^I~,
\\[1ex] 
-\gD\! f_i &= \gk_{ijk}\, H_j\, H_k + \sfrac 1{12}\, c_{2i}~, 
\\[1ex] 
-\gD_i &= - \sfrac 1{24}\, \big( \gk_{ijk}\, V_j \cdot V_k + c_{2i}\big)~. 
}
\end{subequations} 

\subsection{Elements of the effective \texorpdfstring{SO(16)$\times$SO(16)}{SO(16) x SO(16)} theory in four dimensions} 

At leading order in $g_s$ and $\ga'$ we may still employ the N=1 superspace formalism to characterized the bosonic and fermionic fields obtained from a Calabi-Yau compactification of the non-supersymmetric SO(16)$\times$SO(16) theory. The superfields now only include the bosonic or fermionic components present in the non-super\-symmetric theories. Therefore, we briefly indicate the non-vanishing dynamical components of the relevant superfields. This approach is similar to the spurion superfield formalism to encode soft supersymmetry breaking.

\subsubsection*{Tree level superfield action for bosonic fields}

In detail we define the following superfields to describe bosonic moduli and matter scalar fields: 
\equ{
S\big| = \frac 1{2\pi}\,  \Big[e^{-2\gvf} + i\, \gb_0\Big]~, 
\qquad 
T_i\big| =  \frac 1{2\pi} \Big[ - \frac{a_i}{\ell_s^2} + i\, \gb_i \Big]~, 
\qquad 
U_p| = \frac 1{2\pi}\, u_p~, 
\qquad 
Z_m| = z_m~, 
}
as in e.g.~\eqref{S_superfields} and \eqref{T_superfields} with auxiliary field components but without fermionic components. Similarly the vector multiplet $\cV$ does not have any fermionic component: 
\equ{
\sfrac 12\, \big[\,\overline{D}_\dga, D_\ga \big] \cV\big| = 
\gs^\gm_{\dga\ga}\, a_\gm~, 
}
as in~\eqref{S_superfields}. 
The tree-level action for the scalar moduli and matter fields is obtained from the {\em bosonic} K\"ahler potential  
\equ{ \label{BosKahler} 
\cK_\text{bos} = 
- \gk_4^{-2} \, \ln \Big[ S + \overline S \Big]
- \gk_4^{-2} \, \ln \int_X \sfrac 16\, \cJ^3 
- \gk_4^{-2} \, \ln \int_X \bgO\, \gO 
- \cK_\text{bos}^{\um m}\, \bZ_\um e^{2\, q\cdot \cV} Z_m~, 
}
with  
\equ{ 
\cJ = - \sfrac 12\, \Big( T_i + \overline{T}_i \Big)\, \widehat D_i~. 
} 
The tree-level gauge kinetic action is given by the familiar expression 
\equ{ 
S_\text{YM} = 
\frac 14 \int d^4d^2\gth\, \textsc{tr} \Big[ S\, W^2 \Big] + \text{c.c.}~. 
}

\subsubsection*{Tree level superfield action for the fermionic fields}

In addition, the compactification leads to a set of chiral fermions $\gps_f$. We also collect them in chiral superfields $Z_f$, such that their only non-vanishing component is given by 
\equ{
\frac 1{\sqrt 2}\, D_\ga Z_f| = \gps_{f\, \ga}~. 
}
The tree-level kinetic terms of the chiral fermions $\gps_f$ can be encoded in the {\em fermionic} K\"ahler potential 
\equ{ 
K_\text{ferm} = \cK_\text{ferm}^{\undr{f} f}\,  \bZ_\undr{f} e^{2\, q\cdot \cV} Z_f~. 
}
As for the bosonic matter fields the form of this K\"ahler potential is difficult to obtain for general compactifications.

\subsubsection*{One-loop induced anomalous axion-gauge couplings}

We expect that at the one-loop level one encounters corrections that do not respect the relations that rely on supersymmetry. However, the axion couplings that result directly from the Green-Schwarz mechanism in ten dimensions by reduction can still be computed without further difficulties. These couplings are very important as they provide us with detailed anomaly cancellation checks on the fermionic spectra, as in the case of compactifications of the supersymmetric theories. 

The coupling of the axions $\gb_0$ and $\gb_i$ to the four-dimensional gauge fields is determined by the reduction of the various terms in the ten-dimensional Green-Schwarz action. The anomalous gauge transformations of the axions yield 
\equ{ 
\gd \gb_0 =  Q_I\, \ga_I~, 
\qquad 
\gd \gb_i =  Q_i^I\, \ga_I~, 
}
where $\ga_I$ are the Abelian gauge parameters, such that 
$\gd A_{I\,\gm} = - (2\pi)\, \der_\gm \ga_I$. 
The charges $Q_I$ are defined as in~\eqref{defQcharge} and~\eqref{defQiI}. By evaluating the integrals in the SO(16)$\times$SO(16) case we find 
\begin{subequations}
\equa{  \label{qSO16} 
Q_i^{I^\prime} &= V^{\prime I}_i~, \qquad 
Q_i^{I^{\prime\prime}} = V^{{\prime\prime} I}_i~, 
\\[1ex] 
Q^{I^{\prime}} &= \sfrac 1{6}\, \gk_{ijk}\, 
\Big[ 
V_i^{\prime I}\, \Big(V_j^{\prime} \cdot V_k^{\prime}
 - \sfrac 1{2}\, V_j^{\prime\prime}\cdot V_k^{\prime\prime} \Big)
 -\textsc{tr}(H^\prime_i H^\prime_j H^\prime_k H^\prime_I) 
\Big]~,  
\\[1ex]  \non 
Q^{I^{\prime\prime}} &= 
\sfrac 1{6}\, \gk_{ijk}\, 
\Big[ 
V_i^{\prime\prime I}\, \Big(V_j^{\prime\prime} \cdot V_k^{\prime\prime}
 - \sfrac 1{2}\, V_j^{\prime}\cdot V_k^{\prime} \Big)
 -\textsc{tr}(H^{\prime\prime}_i H^{\prime\prime}_j H^{\prime\prime}_k H^{\prime\prime}_I) 
\Big]  
}
\end{subequations} 
The anomalous gauge transformations of the axions lead to a mixing of the axions with the longitudinal parts of the gauge fields and thereby result in massive U(1)s. 
The anomalous couplings of the axions take the form 
\equ{ 
S^\text{GS}_\text{axions} 
\supset \frac 1{2\gp} \int 
\gb_0\, 
\Big\{
\tr F_2^{\prime\,2} + \tr F_2^{\prime\prime\,2}
\Big\} + 
\gb_i\, 
\Big\{ 
\tr \Big[\gD\! f_i^\prime\, F^{\prime\, 2}_2 + \gD\! f_i^{\prime\prime}\, F^{\prime\prime\, 2}_2 \Big]
+ \gD\! f_i^{IJ}\, F^{I}_2 F^J_2 
+ \gD_i \, \textsc{tr}(R_2^2)
\Big\}~. 
}
The coefficients $\gD\! f_i$ are determined from the reduction of the Green-Schwarz term \eqref{GSaction}, 
\equ{ 
S_\text{GS} \supset 
\frac {1}{24(2\gp)^3}\, 
\int \gb_i\, D_i\, X_{4,4}~, 
}
and are given by 
\begin{subequations} 
\equa{
\gD\! f_i^{\prime}~~ &=  \gk_{ijk}\, 
\Big(
\sfrac 1{6}\, V_j^{\prime} \cdot V_k^{\prime}
 - \sfrac 1{12}\, V_j^{\prime\prime} \cdot V_k^{\prime\prime}  
  - H_j^{\prime} H_k^{\prime} \Big)~,  
\\[1ex]  \non 
\gD\! f_i^{\prime\prime}~\, &=   \gk_{ijk}\, 
\Big(
\sfrac 1{6}\, V_j^{\prime\prime} \cdot V_k^{\prime\prime}
 - \sfrac 1{12}\, V_j^{\prime} \cdot V_k^{\prime}  
  - H_j^{\prime\prime} H_k^{\prime\prime} \Big)~, 
\\[1ex]
\gD\! f_i^{J^{\prime}K^{\prime}} & =  
 \sfrac 23\, \gk_{ijk}\, V_{j}^{\prime J} \,V_k^{\prime K}~, 
\qquad 
\gD\! f_i^{J^{\prime\prime}K^{\prime\prime}}  =  
 \sfrac 23\, \gk_{ijk}\, V_j^{\prime\prime J } \, V_k^{\prime\prime K}~,  
\qquad 
\gD\! f_i^{J^{\prime}K^{\prime\prime}} =  
- \sfrac 1{3}\, \gk_{ijk}\, V_j^{\prime J} V_k^{\prime\prime K}~,   
\\[1ex] 
\gD_i &= 0~. 
}
\end{subequations}

\newpage

\begin{table}[t!]
\centering
\renewcommand{\arraystretch}{1.3}
\scalebox{1}{
\begin{tabular}{|cp{4mm}||c||c|}
\hline 
&&{\bf Massless chiral fermions } & \cellcolor{lightgray}{\bf Massless complex bosons}
\\ \hline\hline 
\multirow{3}{*}{\rotatebox{90}{\bf observable}} &  \multirow{2}{*}{\rotatebox{90}{\bf chiral\!\!\!\!\!\!}} &
$
\arry{c}{
    8 (\crep{10};  \rep{1},  \rep{1},  \rep{1}) +
    2 ( \rep{10};  \rep{1},  \rep{1},  \rep{1})
\\[0ex]
 + 24 (  \rep{5};  \rep{1},  \rep{1},  \rep{1}) +
  18 ( \crep{5};  \rep{1},  \rep{1},  \rep{1})
}
$
& \cellcolor{lightgray}
$ 
\arry{c}{ 
   16 (  \rep{5};  \rep{1},  \rep{1},  \rep{1})
\\[0ex]
}$
\\[0ex] 
\hhline{~-||-||-|}
&\vspace{-7mm}\multirow{1}{*}{\rotatebox{90}{\bf \,non-}~\rotatebox{90}{\bf chiral\,}} & \rule{0cm}{7mm}--- & \cellcolor{lightgray} ---
\\[2ex] \hline\hline 
\multirow{3}{*}{\rotatebox{90}{\bf hidden}} &  \multirow{2}{*}{\rotatebox{90}{\bf chiral\!\!\!\!\!\!}} &\!\!\!\!\!\!\!
$
\arry{c}{
  24 (  \rep{1}; \crep{3},  \rep{1},  \rep{1})+
  20 (  \rep{1};  \rep{3},  \rep{1},  \rep{1})+
   2 (  \rep{1};  \rep{3},  \rep{2},  \rep{1})
\\[0ex] 
 +  34 (  \rep{1};  \rep{1},  \rep{2},  \rep{1})+
    28 (  \rep{1};  \rep{1},  \rep{1},  \rep{2})+
   150 (  \rep{1};  \rep{1},  \rep{1},  \rep{1}) 
} 
$\!\!\!\!\!
& \cellcolor{lightgray} \!\!\!\!\!\!\!\!
$ 
\arry{c}{ 
   16 (  \rep{1};  \rep{3},  \rep{1},  \rep{1})+
   12 (  \rep{1}; \crep{3},  \rep{1},  \rep{1}) +
    2 (  \rep{1}; \crep{3},  \rep{2},  \rep{2})
\\[0ex] 
+    4 (  \rep{1};  \rep{1},  \rep{2},  \rep{2})+
    80 (  \rep{1};  \rep{1},  \rep{1},  \rep{1})
 }
 $\!\!\!\!\!\!
\\ 
\hhline{~-||-||-|}
&\vspace{-7mm}\multirow{1}{*}{\rotatebox{90}{\bf \,non-}~\rotatebox{90}{\bf chiral\,}} & \rule{0cm}{7mm}--- & 
$  \cellcolor{lightgray} 5 (  \rep{1};  \rep{3},  \rep{1},  \rep{1})+ 5 (  \rep{1}; \crep{3},  \rep{1},  \rep{1}) $
\\[2ex] \hline
\end{tabular}}
\renewcommand{\arraystretch}{1}
\caption{\label{tb:6genSU5GUT}
This table gives the full charged spectrum of a six generation non-supersymmetric SU(5) GUT theory on the geometry CICY 7862. The final rows in the observable and hidden sectors displays vector-like states which are not detected by the multiplicity operator. 
In fact the charged chiral and full spectrum are identical up to the five $\rep{3}$--$\crep{3}$ pairs in the final row.}
\end{table}

\section{Example of a smooth SM-like model}
\label{sc:SMModel}

In this section we discuss an example to illustrate that it is possible to construct semi-realistic models by compactifying the non-supersymmetric SO(16)$\times$SO(16) theory on smooth Calabi-Yau manifolds. Concretely we consider a line bundle model on the tetra-quadric, i.e.\ on the CICY 7862 geometry. The relevant topological data for this manifold, i.e.\ the intersection numbers and second Chern classes, are given by 
\equ{
\gk_{ijk} = 2~, 
\qquad 
c_{2i} = 24~, 
}
for mutually distinct $i,j,k$ between 1 and 4. On this geometry we can construct a six generation non-supersymmetric SU(5) GUT theory by considering the following line bundle background: 
\equ{ 
\arry{lcl}{
 V_1 &=&  \big(  \sm1,   ~\,1,  ~\, 2,    \sm1,    \sm1,    \sm1,  ~\, 2,  ~\, 1\,)(   
  \sm1,  ~\, 0,    \sm1,  ~\, 0,  ~\, 0,   ~\,0,  ~\, 0,  ~\, 0 \big)~, 
 \\[.5ex] 
 V_2 &=&  \big(  ~\, 0,    \sm1,    \sm1,  ~\, 0,   ~\,0,   ~\,0,  ~\, 0,   ~\,0\,)(  ~\, 1,  ~\, 0,  ~\, 0,    \sm1,  ~\, 0,  ~\, 0,    \sm1,  ~\, 1 \big)~, 
 \\[.5ex] 
 V_3 &=&  \big(  ~\, 0,  ~\, 1,  ~\, 1,  ~\, 0,  ~\, 0,  ~\, 0,    \sm2,  ~\, 0\,)(  ~\, 0,  ~\, 0,    \sm1,  ~\, 2,  ~\, 1,  ~\, 0,  ~\, 2,    \sm2 \big)~, 
 \\[.5ex] 
  V_4 &=&  \big(  ~\,1,  ~\, 0,    \sm1,  ~\, 1,  ~\, 1,  ~\, 1,  ~\, 0,    \sm1\,)(    \sm1,  ~\, 0,  ~\, 2,  ~\, 0,    \sm1,  ~\, 0,  ~\, 0,  ~\, 0 \big)~. 
}
}
The resulting observable and hidden gauge groups are $G_\text{obs} =$ SU(5) and $G_\text{hid} =$ SU(3)$\times$SU(2)$\times$SU(2), respectively. This model satisfies the tree-level DUY equations deep inside the K\"ahler cone: Indeed, if we take the volumes of the four divisors to be related as, 
\equ{
\text{Vol}(D_1)  = \sfrac 12\, \text{Vol}(D_2) = \text{Vol}(D_3) = \text{Vol}(D_4)~, 
}
the tree-level DUY equations are satisfied. 

The full spectrum of a non-supersymmetric six-generation SU(5) GUT model with this line bundle background on the geometry at hand is given in Table~\ref{tb:6genSU5GUT}.  It contains the chiral spectrum computed using the multiplicity operator evaluated on the various ten-dimensional states.  This model contains vector-like fermionic and bosonic exotics at the chiral level. The last rows in the observable and hidden sectors display the additional non-chiral states that can only be determined by cohomology methods reviewed in Subsection~\ref{sc:BeyondChiral}. We see that the number of states that the multiplicity operator misses is very small in this concrete example. 

By a freely acting $\Intr_2$ Wilson line the model becomes a three generation SM-like theory. The Wilson line,  
\equ{
W = \big( 
~\, \sfrac 12,  ~\, 0,  ~\, 0,  ~\, \sfrac 12,  ~\, 0,  ~\,0,  ~\, 0,~\, 0\,)(  ~\,0,  0,~\,   0,~\,   0,~\,   0,~\, 0, ~\,  0,~\, 0
\big)~, 
}
breaks the observable gauge group to $G_\text{obs}=$ SU(2)$\times$SU(3)$\times$U(1)$_Y$. Table~\ref{tb:SMlike} gives the full spectrum in the downstairs description. Again, the final rows in the observable and hidden sectors give the non-chiral states which the multiplicity operator does not see. We recognize that this model is only SM-like and not a true SM candidate: Its spectrum contains eight scalar Higgs doublets, which are all accompanied by scalar color triplets.

\begin{table}[t]
\centering
\renewcommand{\arraystretch}{1.3}
\scalebox{1}{
\begin{tabular}{|lp{4mm}||c||c|}
\hline 
&& {\bf Massless chiral fermions } & \cellcolor{lightgray}{\bf Massless complex bosons}
\\ \hline\hline 
\multirow{2}{*}{\rotatebox{90}{\bf observable}} & \multirow{1}{*}{\rotatebox{90}{\bf chiral\!\!\!}} &\!\!\!\!\!\!
$
\arry{c}{
    4 ( \crep{3},  \rep{2};  \rep{1},  \rep{1},  \rep{1}) + 
     (  \rep{3},  \rep{2};  \rep{1},  \rep{1},  \rep{1})  +    21 (  \rep{1},  \rep{2};  \rep{1},  \rep{1},  \rep{1}) 
 \\[0ex]
 +  16 (  \rep{3},  \rep{1};  \rep{1},  \rep{1},  \rep{1}) + 
   10 ( \crep{3},  \rep{1};  \rep{1},  \rep{1},  \rep{1})  }
$ 
& \cellcolor{lightgray}
 $ 
\arry{c}{ 
    8 (  \rep{3},  \rep{1};  \rep{1},  \rep{1},  \rep{1}) +  
    8 (  \rep{1},  \rep{2};  \rep{1},  \rep{1},  \rep{1}) \\[0ex] 
 } $   \!\!\!\!\!\!\!\!\!\! 
\\[0ex] 
\hhline{~-||-||-|}
&\vspace{-7mm}\multirow{1}{*}{\rotatebox{90}{\bf \,non-}~\rotatebox{90}{\bf chiral\,}} & \rule{0cm}{7mm}--- & \cellcolor{lightgray} ---
\\[2ex] \hline\hline 
\multirow{2}{*}{\rotatebox{90}{\bf hidden~~}} & \multirow{1}{*}{\rotatebox{90}{\bf chiral\!\!\!}} & \!\!\!\!\!\!\!\!\!\!
$ \arry{c}{
   12 (   \rep{1}; \crep{3},  \rep{1},  \rep{1}) +
   10 (  \ \rep{1};  \rep{3},  \rep{1},  \rep{1}) +17 (  \rep{1},  \rep{1};  \rep{1},  \rep{2}, \rep{1})  \\[0ex] 
     + (  \rep{1};  \rep{3},  \rep{2},  \rep{1}) +   
   14 (   \rep{1};  \rep{1},  \rep{1},  \rep{2}) + 
   80 (   \rep{1};  \rep{1},  \rep{1},  \rep{1}) 
} \!\!\!\!\!\!\!\!\!\!
$
&  \cellcolor{lightgray}\!\!\!\!\!\!
 $ 
\arry{c}{ 
     (  \rep{1}; \crep{3},  \rep{2},  \rep{2}) + 
    8 (  \rep{1};  \rep{3},  \rep{1},  \rep{1}) + 6 (  \rep{1}; \crep{3},  \rep{1},  \rep{1})  \\[0ex]   
    +   2 (  \rep{1};  \rep{1},  \rep{2},  \rep{2}) 
   + 40 (  \rep{1};  \rep{1},  \rep{1},  \rep{1})
 }\!\!\!\!\!
 $
\\ 
\hhline{~-||-||-|}
&\vspace{-7mm}\multirow{1}{*}{\rotatebox{90}{\bf \,non-}~\rotatebox{90}{\bf chiral\,}} & \rule{0cm}{7mm}--- & \cellcolor{lightgray}
$   2 (  \rep{1};  \rep{3},  \rep{1},  \rep{1})+ 2 (  \rep{1}; \crep{3},  \rep{1},  \rep{1}) $
\\[2ex] \hline 
\end{tabular}}
\renewcommand{\arraystretch}{1}
\caption{\label{tb:SMlike} 
This table gives the full charged spectrum of an illustrative non-supersymmetric SM-like model on the geometry CICY 7862. In this model the doublet-triplet splitting problem in the scalar Higgs sector is not resolved. 
}
\end{table}

\section{Smooth SM-like models from the standard embedding}
\label{sc:StandardEmbedding}

As observed in \cite{Font:2002pq} and \cite{Blaszczyk:2014qoa} the standard embedding for the SO$(16)\times$SO$(16)$ string on any Calabi-Yau $X$ yields an SO(10) GUT-like theory.
In particular, we have a net number of $h_{21}-h_{11}=\sfrac12 \chi(X)$ fermionic \textbf{16}-plet generations, where $\chi(X)$ is the Euler number of the underlying Calabi-Yau manifold $X$. 
Via a Wilson line associated to a freely acting symmetry $\gG$ there is the possibility to break SO(10) down to the SM gauge group with an additional U(1)$_\text{B-L}$ factor (since a breaking with such discrete Abelian symmetries is always rank-preserving~\cite{Witten1985:StandardEmbedding}) and reduce the number of chiral generations to three. For this reason we have to look for a smooth manifold $X$ that satisfies
\equ{\label{EulerNumberCondition}
\sfrac12 \chi(X/\gG) =\frac{\chi(X)}{2 |\gG|}  = \frac{h_{21}-h_{11}}{|\gG|} 
\stackrel{!}{=} 3\,, 
}
for one of its given freely acting symmetries $\gG$.  

\begin{figure}[t]
\centering
\includegraphics[width=.4\textwidth]{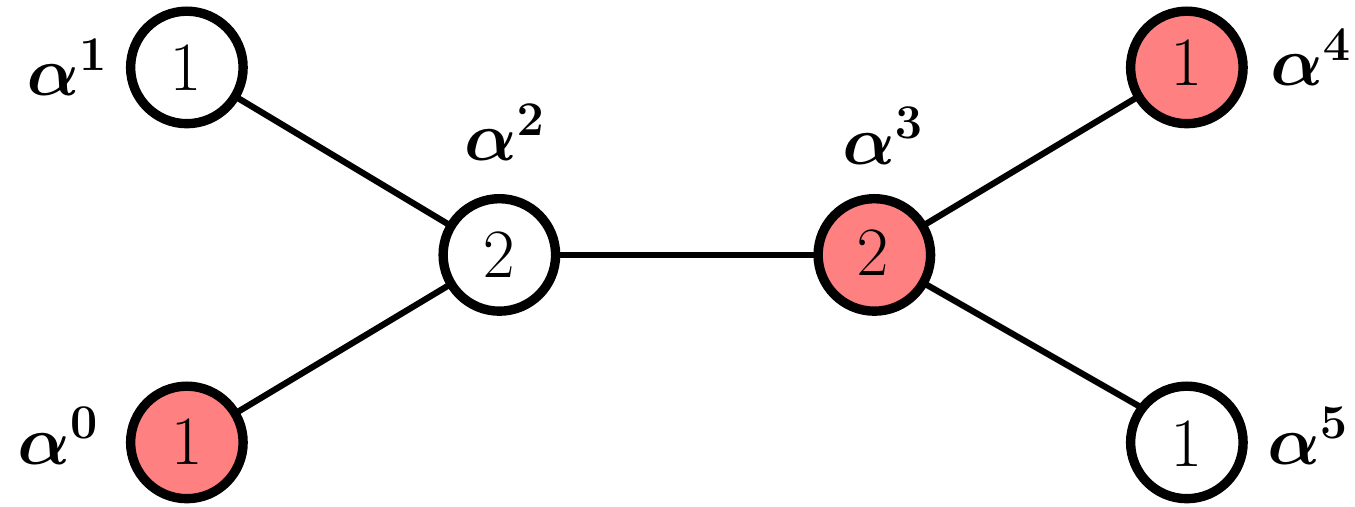}
\caption{\label{fg:DynkinDiagram_SO10}
This figure shows the extended Dynkin diagram of SO(10). For the five simple roots and the extended root the Coxeter labels are given inside the nodes. A breaking to SU$(3)\times$SU$(2)$ is suggested with the colored roots being projected out. All other possible breaking patterns can be obtained by automorphisms of the extended diagram.
} 
\end{figure}

In order to break SO(10) down to $G_\text{obs}=$SU$(3)\times$SU(2)$\times$U(1)$_Y\times$U(1)$_\text{B-L}$, we need at least an freely acting Abelian $\gG = \mathbb{Z}_N$ symmetry with  $N \geq 4$.  To see this, we have depicted the extended Dynkin diagram of the GUT group SO(10) in Figure~\ref{fg:DynkinDiagram_SO10}. Here $-\ga^0=\theta$ denotes the highest root $\theta=a_i\, \ga^i$\,. The Coxeter labels (or marks) $a_i$ for the five simple roots of SO(10) are given inside the nodes; for $\ga^0$ we define $a_0=1$. To determine the unbroken SO(10) simple roots, when modding out an $\mathbb{Z}_N$ Wilson line, we use Dynkin's procedure, as explained in~\cite{Choi2003:DynkinBreaking} to find the unbroken roots:
\equ {\label{DynkinBreaking}
N = s_0 + s_1 + 2 (s_2 + s_3) + s_4 + s_5\,,
}
where $0 \leq s_i \leq N$ for all $i=0,\ldots,5$\,. If $s_i \neq 0$ the corresponding simple root (and any Weyl reflected root) is broken. Using equation~\eqref{DynkinBreaking} we readily compute the lowest order $N$ we need to trigger symmetry breaking down to the gauge group $G_{SM}$ and find the bound $N\geq 4$\,.

Going through the list of classified CICYs and their freely acting symmetries~\cite{Candelas:1987kf,Braun:2010vc}, we find two CICYs with property \eqref{EulerNumberCondition}: CICY~7246 and CICY~7300. Both geometries have $h_{11}=8$, $h_{21}=44$ and can support a $\mathbb{Z}_{12}$ Wilson line. If we allow for an additional ``hidden'' SU(2) symmetry, i.e.\ obtain the downstairs gauge group  $G_\text{obs} \times$SU$(2)$\,, then following \eqref{DynkinBreaking} for the extended Dynkin diagram in Figure~\ref{fg:DynkinDiagram_SO10} we can relax the condition for the order of the Wilson line to $N \geq 3$\,. In the aforementioned list~\cite{Candelas:1987kf,Braun:2010vc} there is only one additional geometry satisfying \eqref{EulerNumberCondition} with $N=3$: CICY~536\,.

\section{Heterotic five-branes in the \texorpdfstring{SO(16)$\times$SO(16)}{SO(16) x SO(16)} theory}
\label{sc:FiveBranes}

One may consider the heterotic string in a background of NS5-branes. The properties of NS5-branes strongly depends on the heterotic theory in question: For the SO(32) and E$_8\times$E$_8$ their properties have been discussed in the literature, see e.g.~\cite{Honecker:2006dt}, while for the SO(16)$\times$SO(16) they are unknown as far as we know. 

To establish the properties of NS5-branes within the non-supersymmetric heterotic string, we will make use of the observation that the perturbative spectrum of the theory can be obtained by non-supersymmetric projections of the SO(32) and E$_8\times$E$_8$ string. As noted above, the full massless spectrum of the SO(16)$\times$SO(16) involves both untwisted and twisted states, if one starts either from the SO(32) or E$_8\times$E$_8$ theory. However, twisted states in one construction are untwisted in the other and vice versa. Hence, by combining the simple orbifold projections within both theories, one has access to the full massless spectrum. Because of this we hope that we can assume the same in the presence of NS5-branes: Hence we will assume that all NS5-brane states in the SO(16)$\times$SO(16) theory can be understood from the supersymmetry-breaking twist acting on the NS5-branes in the SO(32) and E$_8\times$E$_8$ theories. Their anomaly contributions were discussed in detail in~\cite{Honecker:2006dt}; to determine the SO(16)$\times$SO(16) NS5-branes we take inspiration from that discussion. 
The action of this twist is not completely determined by its action on the perturbative parts of these theories. However, as we will see, anomaly cancellation essentially fixes a unique choice.

\subsection*{Perturbative anomaly contributions}

To explain this in detail, we consider a compactification of the SO(16)$\times$SO(16) theory on a smooth K3 with line bundles. The K3 can be characterized by divisors $D_i$ with intersection $\gk_{ij} = D_i D_j$ and $c_1(K3)=0,\ c_2(K3) = 24$. The perturbative anomaly in six dimensions takes the form
\equ{ \label{PertAnom6D} 
I_8^\text{(full)\, pert} =  \int I_{12}^\text{pert} 
= \int  -\frac {1}{24}\, 
\Big( X_{4,0}\, X_{4,4} + X_{2,2}\, X_{6,2} + X_{0,4}\,X_{8,0}
\Big)~, 
}
where the reductions of $X_4$ and $X_8$ read 
\equa{ \non 
X_{4,0}& = \textsc{tr} F_2^{\prime\, 2} + \textsc{tr} F_2^{\prime\prime\, 2} - \textsc{tr} R_2^2~,
\qquad 
X_{2,2} = 2\,\textsc{tr} F_2^{\prime}\cF_2^{\prime} + 2\, \textsc{tr} F_2^{\prime\prime}\cF_2^{\prime\prime}~, 
\\[1ex] 
X_{0,4} &= 2(\tr \cF_2^{\prime\, 2} + \tr \cF_2^{\prime\prime\, 2} - \tr \cR_2^2)
}
and 
\begin{subequations} 
\equa{
X_{4,4} = & 
 -6\, \textsc{tr} F_2^{\prime\, 2} \cF_2^{\prime\, 2} 
 -6\, \textsc{tr} F_2^{\prime\prime\, 2} \cF_2^{\prime\prime\, 2}
 + \textsc{tr} F_2^{\prime\, 2}\, \tr \cF_2^{\prime\, 2} 
 + \textsc{tr} F_2^{\prime\prime\,2}\, \tr \cF_2^{\prime\prime\, 2}
 + 4\, (\tr F_2^{\prime} \cF_2^{\prime})^2  
 + 4\, (\tr F_2^{\prime\prime}\cF_2^{\prime\prime})^2 + 
 \non \\[1ex] & %
- \sfrac 12\,  \textsc{tr} F_2^{\prime\,2}\, \tr \cF_2^{\prime\prime\, 2} 
- \sfrac 12\,  \textsc{tr} F_2^{\prime\prime\, 2}\, \tr \cF_2^{\prime\, 2} 
- 4\, \tr (F_2^{\prime}\cF_2^{\prime})\,  \tr (F_2^{\prime\prime}\cF_2^{\prime\prime})~, 
\\[1ex]  
X_{6,2} = &
 -4\, \textsc{tr} \cF_2^{\prime} F_2^{\prime\, 3} -4\, \textsc{tr} \cF_2^{\prime\prime} F_2^{\prime\prime\, 3} 
 + 2\, \tr (F_2^{\prime} \cF_2^{\prime})\, \textsc{tr} F_2^{\prime\, 2} 
 + 2\, \tr (F_2^{\prime\prime}\cF_2^{\prime\prime})\, \textsc{tr} F_2^{\prime\prime\, 2} + 
 \non \\[1ex]   & 
-  \tr (F_2^{\prime}\cF_2^{\prime})\, \textsc{tr} F_2^{\prime\prime\, 2} 
-  \tr (F_2^{\prime\prime}\cF_2^{\prime\prime})\, \textsc{tr} F_2^{\prime\, 2}~, 
\\[1ex]  \label{X80}
X_{8,0} = & 
- \textsc{tr} F_2^{\prime\, 4}  - \textsc{tr} F_2^{\prime\prime\, 4}  
+\sfrac 14\,  (\textsc{tr} F_2^{\prime\,2})^2
+ \sfrac 14\,  (\textsc{tr} F_2^{\prime\prime\, 2})^2 
- \sfrac 14\,  \textsc{tr} F_2^{\prime\,2}\, \textsc{tr} F_2^{\prime\prime\, 2}~.
}
\end{subequations} 
The first two contributions in~\eqref{PertAnom6D} are automatically cancelled by the reduction of the perturbative Green-Schwarz mechanism~\eqref{GSaction}; therefore we will not consider those contributions further here. For the third and final contribution this is not the case, since the integrated Bianchi identity, the integral over $X_{0,4}$, defines the five-brane charge 
\equ{ 
N =  \frac12 \int X_{0,4} = 
\int (\tr \cF_2^{\prime\, 2} + \tr \cF_2^{\prime\prime\, 2} - \tr \cR_2^2)
= \gk_{ij}\, V_i\cdot V_j + 48~. 
}
We see that for $N\neq 0$ the perturbative part of the SO(16)$\times$SO(16) suffers from irreducible anomalies of SO(16) or appropriate subgroups thereof, but not from an irreducible gravitational anomaly: 
\equ{ 
I^\text{pert}_8 = 
-\frac {2\,N}{24}\,  \Big\{  
- \textsc{tr} F_2^{\prime\, 4}  - \textsc{tr} F_2^{\prime\prime\, 4}  
+\sfrac 14\,  (\textsc{tr} F_2^{\prime\,2})^2
+ \sfrac 14\,  (\textsc{tr} F_2^{\prime\prime\, 2})^2 
- \sfrac 14\,  \textsc{tr} F_2^{\prime\,2}\, \textsc{tr} F_2^{\prime\prime\, 2}
\Big\}~. 
} 
(Here we only give the anomaly contributions that are not cancelled by the reduction of the ten-dimensional Green-Schwarz mechanism.) 
This means that the NS5-branes in the SO(16)$\times$SO(16) theory have to cancel these irreducible gauge anomalies, but  their irreducible gravitational and non-perturbative gauge anomalies all have to vanish independently.

\subsection*{SO(32) NS5-brane anomaly contributions}

\begin{table}
\[
\begin{array}{|c|c|c|}
\hline 
\multicolumn{3}{|c|}{\text{\bf SO(32) NS5-branes}} 
\\ \hline\hline
\text{Sp}(2\tN)~\text{6D} & \text{Bi-fundamental} & \text{Anti-symmetric} 
\\
 \text{vector multiplet}  & \text{half-hypermultiplets} & \text{hypermultiplets} 
 \\ \hline
{\textsf{V}} \qquad ([2\tN]_2^+)_+ & {\textsf{H}} \qquad (32; 2\tN)_-& {\textsf{C}} \qquad  ([2\tN]_2^-)_-  
\\ \hline \hline 
\cellcolor{lightgray}
\text{Sp}(2\tN')\times \text{Sp}(2\tN'') & 
\cellcolor{lightgray} 
(16'; 2\tN'') + (16''; 2\tN') & 
\cellcolor{lightgray}
(2\tN', 2\tN'')  
\\[-.5ex] 
\cellcolor{lightgray}
\text{gauge fields} & 
\cellcolor{lightgray}
\text{scalars} & 
\cellcolor{lightgray}
\text{scalars} 
\\ \hline 
(2\tN', 2\tN'')_+ & 
\sfrac 12 (16'; 2\tN')_- + \sfrac 12 (16''; 2\tN'')_- & 
([2\tN']_2^-)_- + ([2\tN'']_2^-)_- 
\\[-.5ex] 
\text{gauginos} & \text{half-hyperinos} & \text{hyperinos} 
\\ \hline 
\end{array}
\]
\caption{\label{tb:SONS5branes}
The top part of this table gives the matter spectra on $\tN$ coinciding NS5-branes in the heterotic SO(32) theory. The notation $[2\tN]_2^\pm$ denotes the totally symmetric/anti-symmetric rank-two tensor of Sp(2$\tN$). The subscript $\pm$ on the various representations indicates the six-dimensional chirality of the corresponding fermions. 
(We use the same convention for the perturbative theories in which the chiralities of E$_8\times$E$_8$ and SO(32) are taken to be opposite.) The bottom part of the table displays the remaining states after the non-supersymmetric projection has been performed. 
}
\end{table}

Next, we briefly discuss the spectra on $\tN$ coinciding NS5-branes in the SO(32) theory, cf.\ Table~\ref{tb:SONS5branes}. To figure out how the non-supersymmetric twist acts on the NS5-brane spectrum, we first recall that the action on the SO(32) gauge fields can be represented as 
\equ{
A_M \ra U_\text{SO}\, A_M\, U^T_\text{SO}~, 
\quad\text{where}\quad
U_\text{SO} = \pmtrx{ \Id_{16} & 0 \\ 0 & -\Id_{16} }~.
}
We extend the non-supersymmetric twist on the SO(32) NS5-brane fields
\equ{ \label{NonSusyTwistNP} 
{\textsf{V}} \ra U_\text{Sp}\, {\textsf{V}} \, U^T_\text{Sp}~, 
\quad 
\textsf{H} \ra U_\text{SO}\, \textsf{H}\, U_\text{Sp}^T~, 
\quad 
\textsf{C} \ra U_\text{Sp}\, \textsf{C}\, U_\text{Sp}^T~, 
~~\text{where}~~
U_\text{Sp} = \pmtrx{ \Id_{2\tN'} & 0 \\ 0 & -\Id_{2\tN''} }~, 
}
such that $\tN' + \tN'' = \tN$. 
The choice of the supersymmetry-breaking twist on the NS5-brane states is a priori not unique even up to similarity transformations. In particular, there may be an additional minus sign in the transformation of the anti-symmetric hypermultiplet $\textsf{C}$. However, as we are then not able to cancel the irreducible Sp($2\tN'$) and Sp($2\tN''$) anomalies, we disregard such possibilities. 

The bosonic and fermionic states that survive this supersymmetry-breaking twist~\eqref{NonSusyTwistNP} are given in the last two rows of Table~\ref{tb:SONS5branes}. The fermionic part of the spectrum produces an anomaly 
\equ{ 
\label{eq:I8SO32}
I_8^\text{SO\,NS5} = 
\frac{\tN''-\tN'}{12}
\Big( \textsc{tr} \widetilde{F}_2^{\prime\, 4}  - \textsc{tr} \widetilde{F}_2^{\prime\prime\, 4} \Big)
-\frac{15(\tN' + \tN'') + 2 (\tN' - \tN'')^2}{128}\,\Big(
\frac1{45}\, \textsc{tr} R_2^4 + \frac 1{36}\, (\textsc{tr} R_2^2)^2 
\Big) 
\non \\[1ex] 
+ \frac 1{96}\,  \textsc{tr} R_2^2\, \Big( 
\tN'\, \textsc{tr} F_2^{\prime\, 2} +  \tN''\, \textsc{tr} F_2^{\prime\prime\, 2}    
+(2\,\tN' - 2\,\tN'' + 6)\, \textsc{tr} \widetilde{F}_2^{\prime\, 2} 
+(2\, \tN'' - 2\,\tN' +6)\, \textsc{tr} \widetilde{F}_2^{\prime\prime\, 2}    
\Big)
\non \\[1ex]  
- \frac{\tN'}{24}\, \textsc{tr} F_2^{\prime\, 4} 
- \frac{\tN''}{24}\, \textsc{tr} F_2^{\prime\prime\, 4}   
- \frac 18\,  (\textsc{tr} \widetilde{F}_2^{\prime\, 2})^2 
- \frac 18\,  (\textsc{tr} \widetilde{F}_2^{\prime\prime\, 2})^2 
+ \frac 14\, \textsc{tr} \widetilde{F}_2^{\prime\, 2} \, \textsc{tr} \widetilde{F}_2^{\prime\prime\, 2}
\non \\[1ex] 
- \frac 18\,  \textsc{tr} F_2^{\prime\, 2}\, \textsc{tr} \widetilde{F}_2^{\prime\, 2}
-\frac 18\,  \textsc{tr} F_2^{\prime\prime\, 2}\, \textsc{tr} \widetilde{F}_2^{\prime\prime\, 2} 
~. 
} 
Here $\widetilde{F}_2^{\prime},\ \widetilde{F}_2^{\prime\prime}$ denote the gauge field strengths of the Sp($2\tN^{\prime}$), Sp($2\tN^{\prime\prime}$) groups and $\textsc{tr}$ is the trace in the fundamental of Sp-groups.
The overall sign of the anomaly contributions is fixed by the following consideration: In our convention the ten-dimensional chirality of the SO(32) theory is opposite to that of the E$_8\times$E$_8$ theory. We require that the NS5-branes coming from the SO(32) theory preserve the supersymmetry realized in the compactification of the SO(32) theory.

The anomaly polynomial contains irreducible anomalies of (various subgroups of) SO(16)${}^\prime\times$SO(16)${}^{\prime\prime}$, Sp($2\tN'$) and Sp($2\tN''$) and irreducible gravitational anomalies. The irreducible 
Sp($2\tN'$) and Sp($2\tN''$) anomalies drop out and the irreducible anomalies of (subgroups of) SO(16)${}^\prime\times$SO(16)${}^{\prime\prime}$ cancel those due to the last term in~\eqref{PertAnom6D} (inserting~\eqref{X80}) provided that we choose 
\equ{ \label{CancelIrrAnoms} 
\tN' = \tN'' = 2\,N~. 
}
For this choice the irreducible gravitational anomaly remains.  
It is remarkable that the irreducible parts of SO(16)$\times$SO(16) anomalies in \eqref{eq:I8SO32} which are independent of $\tN',\tN''$ cancel among themselves.

\subsection*{E$_8\times$E$_8$ NS5-brane anomaly contributions}

\begin{table}
\[
\begin{array}{|c|c|}
\hline 
\multicolumn{2}{|c|}{\text{\bf E$\boldsymbol{_8\times}$E$\boldsymbol{_8}$ NS5-branes}} 
\\ \hline\hline
 \text{Tensor multiplets} & \text{Hypermultiplets} 
\\ \hline
 {\textsf{T}}_s &
 {\textsf{H}}_s~,~s=1,\ldots, \tn
\\ \hline \hline 
\cellcolor{lightgray}
\tn'~\text{anti-self-dual tensors} & 
\cellcolor{lightgray}
\tn''~\text{complex scalars} 
\\ \hline 
 \tn''~(\text{tensorinos})_+
  & \tn'~(\text{hyperinos})_+ 
 \\ \hline 
\end{array}
\]
\caption{\label{tb:E8xE8NS5branes}
The top part of this table gives the matter spectra on $\tn$ coinciding NS5-branes in the E$_8\times$E$_8$ theory. 
The bottom part of the table displays the remaining states after the non-supersymmetric projection has been performed. 
}
\end{table}

We see that the non-supersymmetric projection of the SO(32) NS5-branes leads to an irreducible gravitational anomaly. To cancel this anomaly we can use the E$_8\times$E$_8$ NS5-branes that support six-dimensional tensor multiplets $\textsf{T}_s$ and hypermultiplets $\textsf{H}_s$, $s=1,\ldots, \tn$. The tensor multiplets include anti-self-dual tensors. The scalars in the hypermultiplets parameterize the positions of the NS5-branes on the K3~\cite{Seiberg:1996vs}. 

As for the SO(32) NS5-branes, we have to decide how the tensor- and hypermultiplet components transform under the supersymmetry-breaking twist. We take: 
\equ{
\textsf{T}_{s'} \ra \textsf{T}_{s'}~,
\quad 
\textsf{T}_{s''} \ra -\textsf{T}_{s''}~,
\qquad 
\textsf{H}_{s''} \ra \textsf{H}_{s''}~,
\quad 
\textsf{H}_{s'} \ra -\textsf{H}_{s'}~,
}
with $s' = 1,\ldots, \tn'$ and $s''=1,\ldots,\tn''$, such that $\tn'+\tn''=\tn$. The surviving spectrum is given in the bottom part of Table~\ref{tb:E8xE8NS5branes}. The resulting gravitational anomaly reads 
\equ{
I_8^\text{E$_8\times$E$_8$\,NS5} =  
\frac{\tn'}{128}\,\Big(
\frac{28}{45}\, \textsc{tr} R_2^4 - \frac 8{36}\, (\textsc{tr} R_2^2)^2 
\Big)
+
\frac{\tn'+\tn''}{128}\,\Big(
\frac1{45}\, \textsc{tr} R_2^4 + \frac 1{36}\, (\textsc{tr} R_2^2)^2 
\Big)~. 
} 
The first contribution comes from the surviving anti-self-dual tensor fields and the second from the surviving hyperinos and tensorinos. Consequently, if we take 
\equ{
\label{eq:E8NS5braneDistribution}
\tn'' = 60\, N - 29\, \tn'~, 
}
we see that all irreducible gravitational anomalies are cancelled.

\subsection*{Factorization}

The remaining reducible anomalies read 
\equa{  \non 
I_8^\text{red} =& 
 \frac{2\,N}{96}\Big[
\textsc{tr} F^{\prime\, 2}\, \textsc{tr} F^{\prime\prime\, 2} 
-  (\textsc{tr} F^{\prime\, 2})^2 
- (\textsc{tr} F^{\prime\prime\, 2})^2 
+ \textsc{tr} R_2^2 \,\big( \textsc{tr} F^{\prime\, 2} + \textsc{tr} F^{\prime\prime\, 2} \big) 
\Big]
- \frac{\tn'}{128}\, (\textsc{tr} R_2^2 )^2
\\[1ex] \label{RedAnomaly}
&- \frac 18\,  \textsc{tr} F^{\prime\, 2} \,\textsc{tr} \tF^{\prime\, 2}
- \frac 18\,  \textsc{tr} F^{\prime\prime\, 2}\, \textsc{tr} \tF^{\prime\prime\, 2}
+ \frac 1{16}\, \textsc{tr} R_2^2\, \Big(\textsc{tr} \tF^{\prime\, 2} + \textsc{tr} \tF^{\prime\prime\, 2} \Big) 
\\[1ex] \non 
&- \frac 18\,  (\textsc{tr} \tF^{\prime\, 2})^2
- \frac 18\,  (\textsc{tr} \tF^{\prime\prime\, 2})^2
+ \frac 14\,  \textsc{tr} \tF^{\prime\, 2}\, \textsc{tr} \tF^{\prime\prime\, 2}~. 
}
This expression is symmetric under the simultaneous exchange of $F_2^{\prime\,2} \leftrightarrow F_2^{\prime\prime\,2}$ and $\tF_2^{\prime\,2} \leftrightarrow \tF_2^{\prime\prime\,2}$. Consequently, the anomaly canceling diagrams need to have the same symmetry. 

The field strength of the anti-symmetric tensors $B^s_2$ are denoted by $H_3^s = d B_2^s+$CS$_3$-terms, such that 
\equ{
d H_3^s = \frac {\ga'}8\, \Big\{
a\,( \textsc{tr} F_2^{\prime\, 2} - \textsc{tr} F_2^{\prime\prime\, 2}) 
+ \ta\, (\textsc{tr} \tF_2^{\prime\, 2} - \textsc{tr} \tF_2^{\prime\prime\, 2})
\Big\} ~. 
}
Here $a, \ta$ are constants to be determined below from anomaly factorization. The relevant part of the six-dimensional NS5-brane action including Green-Schwarz-like Chern-Simons interactions can be represented as 
\equa{ \label{GSonNS5}
S = &\sum_{s=1}^{\tn'}  \int \Big\{ 
-\frac {\pi}{\ell_s^4}\, H^s_3\wedge*H^s_3 
+ \frac{c}{24(2\pi)\ell_s^2}\, B_2^s \Big(
\textsc{tr} F_2^{\prime\, 2} - \textsc{tr} F_2^{\prime\prime\, 2}
 \Big) 
\\[1ex]  \non 
& +\frac{1}{24(2\pi)\ell_s^2}\, B_2 \Big(  
b\, (\textsc{tr} F_2^{\prime\, 2} + \textsc{tr} F_2^{\prime\prime\, 2}) + 
\tb\, (\textsc{tr} \tF_2^{\prime\, 2} + \textsc{tr} \tF_2^{\prime\prime\, 2})
- b_R \, \textsc{tr} R_2^2\Big)
\Big\}~, 
}
where $b, \tb, b_R, c$ are further constants. This leads to the following anomaly contributions:
\begin{subequations}
\equa{ 
I^{(1)}_8 & = 
\frac 1{96}\, 
\Big[ \textsc{tr} F^{\prime\, 2} + \textsc{tr} F^{\prime\prime\, 2} - \textsc{tr} R^{\prime\, 2} \Big] 
\Big[
b\, (\textsc{tr} F_2^{\prime\, 2} + \textsc{tr} F_2^{\prime\prime\, 2}) + 
\tb\, (\textsc{tr} \tF_2^{\prime\, 2} + \textsc{tr} \tF_2^{\prime\prime\, 2})
- b_R \, \textsc{tr} R_2^2
\Big]~, \label{eq:6DAnomPart1}
\\[1ex]
I^{(2)}_8 & = 
\frac{\tn'}{128}\, 
\Big[ 
a\,( \textsc{tr} F_2^{\prime\, 2} - \textsc{tr} F_2^{\prime\prime\, 2}) 
+ \ta\, (\textsc{tr} \tF_2^{\prime\, 2} - \textsc{tr} \tF_2^{\prime\prime\, 2})
\Big]^2~, \label{eq:6DAnomPart2}
\\[1ex] 
I^{(3)}_8 & = 
\frac{\tn' c}{192}\, 
\Big[ 
a\,( \textsc{tr} F_2^{\prime\, 2} - \textsc{tr} F_2^{\prime\prime\, 2}) 
+ \ta\, (\textsc{tr} \tF_2^{\prime\, 2} - \textsc{tr} \tF_2^{\prime\prime\, 2})
\Big] 
\Big[
\textsc{tr} F_2^{\prime\, 2} - \textsc{tr} F_2^{\prime\prime\, 2}
\Big]~. \label{eq:6DAnomPart3}
}
\end{subequations} 
These contributions respect the same permutation symmetries as we observed in~\eqref{RedAnomaly}. The factor $\tn'$ in the second and third contribution arises because there are $\tn'$ tensors $B^s_2$ which mediate the anomaly cancellation. They cancel exactly when the coefficients are chosen as 
\equa{
\label{eq:SO16NS5AnomalyFactorization}
\tn' = 2N~, 
\qquad
b = \sfrac 12\, N~, 
\qquad
\tb = 6~, 
\qquad 
b_R = \sfrac 32\, N~, 
\nonumber \\[1ex]
a = 
\pm\frac{1 \mp \sqrt{1 -  2N}}{\sqrt {2N}}~, 
\quad
 \ta = \frac{4}{\sqrt{2N}} ~, 
\quad
c = 3\, \frac{\sqrt{1-2N}}{\sqrt{2N}}~. 
}
Note that there are two more solutions obtained from inverting the signs of $(a,\ta,c)$ simultaneously which is due to the parameterization in \eqref{eq:6DAnomPart3}. All solutions have $\tn' = 2N$, which means that, using \eqref{CancelIrrAnoms} and \eqref{eq:E8NS5braneDistribution}, $\tN'=\tN''=\tn'=\tn''$, i.e.\ the number of the NS5-branes from SO(32) and E$_8\times$E$_8$ match. 

To summarize we find the rather surprising result that \eqref{eq:SO16NS5AnomalyFactorization} has only one genuine solution which has
\equ{
\tN'=\tN''=\tn'=\tn''=1~, 
\quad\text{and}\quad
N = 1/2~,
}
i.e.\ a single NS5-brane
This result has been obtained under the following assumptions: 
\begin{enumerate}[i)] 
\item We can understand all NS5-brane properties by studying the untwisted sector of the super\-symmetry-breaking twist of the supersymmetric E$_8\times$E$_8$ and SO(32) theory combined. 
\item  The SO(32) and E$_8\times$E$_8$ NS5-branes preserve the same supersymmetry as present in the perturbative sector of SO(32) and E$_8\times$E$_8$ theories, respectively. 
\item We have ignored the possibility of having states that stretch between the E$_8\times$E$_8$ and SO(32)-type NS5-branes.
\item  We made a restrictive ansatz~\eqref{GSonNS5} for the generalized Green-Schwarz couplings on the NS5-branes.
\end{enumerate} 
It would be important to provide either further evidence for this result or to find potential problems and/or generalizations of our arguments. Moreover, we wonder whether we should interpret the five-branes in the SO(16)$\times$SO(16) string as one or two types of NS5-branes. Even more important is the question whether these five-branes could become an additional source for tachyons.

\clearpage
\begin{appendices}
\def\theequation{\thesection.\arabic{equation}} 
\setcounter{equation}{0}

\section{Traces}
\label{sc:Traces} 

A representation $\rep{R} = \{ |p\rangle \}$ is characterized by a set of vectors $|p\rangle$ corresponding to the weights $p\in W_\rep{R} = \{p \}$. We identify a representation module $\rep{R}$ with its weights system $W_\rep{R}$ for notational convenience and write $W_\rep{R} = \rep{R}$. 
Some representations and their weights of SU and SO-groups are indicated in Tables~\ref{SUweights} and~\ref{SOweights}, respectively. The dimensions of a representation $\rep{R}$ is denoted by $|\rep{R}|$. 

The generators $T_A$ of a group $G$ are labeled by $A$. We take the {\em same} Cartan generators, denoted by $H_I$, in all three heterotic theories.
Their eigenvalues are the components of the weights, 
\equ{ 
H_I\, |p\rangle = p_I\, |p\rangle~. 
}
The remaining generators are denoted by $E_\ga$ where $\ga$ are the roots of the group, i.e.\ the weights in the adjoint representation. 

The trace of an operator $A_G$ over a representation $\rep{R}$ is defined as 
\equ{
\tr_\rep{R} (A_G) = \sum_{p\in \rep{R}} \langle p | A_G | p \rangle~. 
}
The subscript $G$ indicates that one is performing the trace of an operator which is a function of objects that are functions of algebra elements associated to the group $G$. 
The character of an operator $A_G$ over a representation $\rep{R}$ is defined as 
\equ{ 
\text{ch}_\rep{R}(A_G) = \tr_\rep{R}\big( e^{A_G} \big)~.  
}
Characters are compatible with direct sums and tensor products in the sense that
\equ{ 
\text{ch}_{\rep{r}\oplus\rep{R}}(A_G) = \text{ch}_\rep{r}(A) +\text{ch}_\rep{R}(A_G)~, 
\qquad 
\text{ch}_{\rep{r}\otimes\rep{R}}(A_G) = \text{ch}_\rep{r}(A_G)\, \text{ch}_\rep{R}(A_G)~. 
}
Consequently, for anti-symmetric tensor products one has: 
\begin{subequations} 
\equa{ 
\text{ch}_\rep{[R]_2^-} (A_G) & = 
\sfrac 12\, \Big( \text{ch}_\rep{R}(A_G)^2 - \text{ch}_\rep{R}(2A_G) \Big)~, 
\\[1ex]  
\text{ch}_\rep{[R]_3^-} (A_G) & = 
\sfrac 16\, \Big( \text{ch}_\rep{R}(A_G)^3 -3\, \text{ch}_\rep{R}(A_G)\, \text{ch}_\rep{R}(2A_G)  + 2\, \text{ch}_\rep{R}(3A_G) 
\Big)~, 
}
\end{subequations} 

\begin{table}[ht]
\[
\arry{|c|c|c|c|c|}{
\hline 
\multicolumn{5}{|c|}{\text{\bf SU($N$) representations}} \\ 
\text{\bf Name}& \rep{R}  & \text{\bf Weights }p & |\rep{R}| & \ell(\rep{R})  
\\ \hline\hline
\text{Vector} & \rep{F} = \rep{N} & \big(\undr{1, 0^{N-1}}\big) & N & 1 
\\
\text{Adjoint} & \rep{Ad} & \big(\undr{\pm1^2, 0^{N-2}}\big) &N^2-1 &2N  
\\
\text{Rank-$2$ tensor} & \rep{[N]^-_2} & \big(1^2, 0^{N-2} \big) &\sfrac 12 N(N-1) & N-2 
\\
\text{Rank-$3$ tensor} & \rep{[N]^-_3} & \big(1^3, 0^{N-3} \big) &\sfrac16N(N-1)(N-2) &\sfrac 12(N-2)(N-3) 
\\
\vdots & \vdots & \vdots & \vdots & \vdots 
\\ \hline 
}
\]
\caption{\label{SUweights} 
Some representations of SU($N$) and their weights are indicated.}
\end{table}

\begin{table}[ht] 
\[
\arry{|c|c|c|c|c|}{
\hline 
\multicolumn{5}{|c|}{\text{\bf SO(2$N$) representations}} \\ 
\text{\bf Name}& \rep{R}  & \text{\bf Weights }p & |\rep{R}| & \ell(\rep{R})  
\\ \hline\hline
\text{Vector} & \rep{2N} & \big(\undr{\pm1, 0^{N-1}}\big) & 2N & 2   \\
\text{Adjoint} & \rep{Ad} & \big(\undr{\pm1^2, 0^{N-2}}\big) & N(2N-1) & 4(N-1)   \\
\text{Spinor}_\pm & \rep{2_\pm^{N-1}} & \big( \pm \sfrac 12^N \big) & 2^{N-1} & 2^{N-3} \\ 
& &  \#(-) = \text{even/odd} & & 
\\ \hline 
}
\]
\caption{\label{SOweights} 
Some representations of SO($2N$) and their weights are indicated.}
\end{table}

We denote by
\equ{ 
\tr (A_\text{SU}) = \tr_\rep{N}(A_\text{SU}) 
}
the trace as if it is the trace over the fundamental representation of an SU group, i.e.\ the vector representation $\rep{N}$ of the SU($N$) group. Similarly, we define for the trace over the fundamental (vector) representations $\rep{2N}$ of SO($2N$) and Sp($2N$), 
\equ{ 
\textsc{tr}(A_\text{SO}) = \tr_\rep{2N}(A_\text{SO})~, 
\qquad 
\textsc{tr}(A_\text{Sp}) = \tr_\rep{2N}(A_\text{Sp})~, 
}
For SO- and Sp-groups this means that 
\equ{
\textsc{tr} (A_\text{SU}) = \tr_\rep{2N} (A_\text{SU}) = 2\, \tr_\rep{N}(A_\text{SU}) = 2\, \tr(A_\text{SU})~, 
}
since we have the branching $\rep{2N} \ra \rep{N} + \rep{\overline{N}}$ when we consider $A_\text{SU}^\dag = A_\text{SU}$ in the SU($N$) subalgebra  of the SO($2N$) algebra. We often write traces of objects valued in a certain algebra as traced over another representation associated to a different algebra. In this case this should be understood as defining some useful notation, not literally as the trace written (as that would not necessarily make sense). For example, the l.h.s.\ of 
\equ{ 
\text{tr}(A_\text{SO}) = \tr_\rep{N}( A_\text{SO}) := \frac 12\, \tr_\rep{2N}(A_\text{SO})~, 
} 
does not make sense because the smallest representation of SO($2N$) is $\rep{2N}$ and not the fundamental representation $N$ of the SU($N$) group. Hence, here the l.h.s.\ is defined as the r.h.s.\ including the normalization factor $1/2$. Similar one often uses the trace of the adjoint of E$_8$ written as the trace over the fundamental of SO($16$): 
\equ{ 
\textsc{tr}(A_\text{E$_8$}) = 
\tr_{16} (A_\text{E$_8$}) := 
\frac 1{30}\, \tr_{248} (A_\text{E$_8$})~. 
}
We reserve the notation $\Tr$ to denote the trace over the full fermionic spectrum. For the supersymmetric E$_8\times$E$_8$ and SO(32) theories, this is then the trace in the adjoint of the respective gauge group. 

The quadratic Casimir operator of the algebra is given by
\equ{
C_G = \sum_A (T_A T_A)_G~.
}
Denoting the eigenvalue of this Casimir evaluated on a representation $\rep{R}$ by $C(\rep{R})$, we have 
\equ{ 
C(\rep{R})\, |\rep{R}| = \sum_A \tr_\rep{R} (T_A T_A)_G~. 
}
This means that 
\equ{ \label{TraceFundSU}
\tr_\rep{N}(T_A T_B)_\text{SU} = \gd_{AB}~, 
\qquad 
\tr_\rep{N}(C) =  |\rep{Ad}|~. 
}
The index $\ell(\rep{R})$ of a representation $\rep{R}$ of a given group $G$ is defined as 
\equ{
\tr_\rep{R} (T_A T_B) = \ell(\rep{R})\, \tr_\rep{N}(T_A T_B)~. 
}
By setting $B=A$ and summing over $A$, we obtain the trace~\eqref{TraceFundSU} on the right-hand side. Because of the weights of the spinor representation $\rep{2^{N-1}_+}$, it branches into a sum of even rank anti-symmetric tensor representations $\rep{[N]_{2k}^-}$. Using this one can determine the index of the spinor representation if the indices of the anti-symmetric tensor representations of the SU group are known. The resulting values for some representations indices of SU and SO-groups can be found in Tables~\ref{SUweights} and~\ref{SOweights}.

For quadratic traces of two Cartan generators in the fundamental representations of the SU and SO groups we find: 
\equ{
\tr_\rep{N} (H_I\, H_J) = \gd_{IJ}~, 
\qquad 
\tr_\rep{2N} (H_I\, H_J) = 2\, \gd_{IJ}~, 
}
respectively, and similarly for quartic traces: 
\equ{ 
\tr_\rep{N} (H_I\, H_J\, H_K\, H_L) = \gd_{IJKL}~, 
\qquad 
\tr_\rep{2N} (H_I\, H_J\,H_K\, H_L) = 2\, \gd_{IJKL}~, 
} 
where $\gd_{IJKL} =1$ when all indices are equal and zero otherwise. 
We should stress that these trace identities only hold when traced over Cartan generators as stated here; generic quartic traces are more complicated.

\def\theequation{\thesection.\arabic{equation}} 
\setcounter{equation}{0}

\section{Line bundle description as \texorpdfstring{S(U(1)$^{\boldsymbol{n+1}}$)}{S(U(1)\^{}{n+1})} bundles}
\label{sc:IdUbundles} 

The authors of some of the literature on line bundles on Calabi-Yau manifolds~\cite{Anderson:2011ns,Anderson:2012yf,Anderson:2013xka} use a different parameterization for the embedding of the structure group of the vector bundle into the primordial gauge group. We briefly review this parameterization in order to facilitate contact with our description. A vector bundle $\mathcal{V}$ with structure group S(U(1)${}^{n+1}$) can be obtained as a direct sum of line bundles
\equ{ 
\mathcal{V} = \bigoplus_{a=1}^{n+1}
 \mathcal{O}(k^{(a)}_1,\ldots, k^{(a)}_{h_{11}})~, 
}
labeled by $n\,h_{11}$ integers $k^{(a)}_i$. This leads to a gauge flux that can be represented as
\equ{
\frac{\mathcal{F}}{2\pi}  = 
 k^{(a)}_i\, D_i \, H_{(a)}~, 
}
where $H_{(a)}$ are the $n+1$ U(1) generators of the Cartan of U(n+1). In particular, $k^{(a)}$ can be identified with the charge of the $a$-th $\mathbf{10}$-plet. Here we have expanded the first Chern classes $c_1(\mathcal{O}(k^{(a)}_1,\ldots k^{(a)}_{h_{11}}) = k^{(a)}_i\, \widehat{D}_i$ associated to the divisors ${D}_i$. In order to ensure that we have an S(U(1)${}^{n+1}$) and not a U(1)${}^{n+1}$ structure group we require that $c_1(\mathcal{V}) =0$, i.e.\ the corresponding gauge flux is traceless: $k^{(n+1)}_i = - k^{(1)}_i - \ldots - k^{(n)}_i$.

\subsection[A line bundle vector representation of \texorpdfstring{S(U(1)$^{5}$}{S(U(1)\^{}5}) bundles]{A line bundle vector representation of \texorpdfstring{S(U(1)$^{\boldsymbol{5}}$}{S(U(1)\^{}5}) bundles}

In the literature mostly S(U(1)$^5$)$ \supset $E$_8$ bundles are discussed. To translate such bundle backgrounds into the language used in this work we observe the following: The E$_8$ gauge group will contain an unbroken SU(5) group if we choose\footnote{Other parameterizations corresponding to different embeddings in the E$_8$ are also possible.} (up to overall permutations)
\equ{ \label{SU5inE8emb} 
V_i = (a_i^5, b_i, c_i, d_i)~, 
}
provided that the coefficients $a_i\neq0, b_i, c_i, d_i$ are sufficiently generic, i.e.\ no entries are equal or opposite and the sums of all entries mod two does not vanish. Using this choice, all unbroken SU(5) roots, $(\undr{\sm1,1,0^3},0^3)(0^8)$, are vectorial. In the non-supersymmetric case, this ansatz leads to the breaking of one of the SO(16) factors to SU(5).

To identify the integers $k^{(a)}_i$ with the quantities appearing in the parameterization $V_i$ given in~\eqref{SU5inE8emb} we compare the value of the charges of the $\mathbf{10}$-plets of SU(5) in both descriptions. The motivation to use the $\mathbf{10}$-plets for this matching is given by the branching the adjoint of E$_8$ 
\equ{ 
\mathbf{248} \ra 
(\rep{24},\rep{1}) + (\rep{1},\rep{24}) + (\rep{10}, \rep{5}) + (\crep{10}, \crep{5})  
+ (\rep{5}, \crep{10}) + (\crep{5}, \rep{10}) 
}
under $\text{E}_8 \ra \text{SU}(5)\times \text{SU}(5)$. Hence we see the that U(1)$^5$ charges under the gauge fluxes supported on the divisor $\widehat{D}_i$ of the $\rep{5}$-components are simply $k_{(a)}^i$. Using this we obtain the following relation between the U(1) bundle charges in both languages: 
\equ{ 
\arry{|c|c|c|}{ 
\hline 
\mathbf{10}\text{\bf -plets} & V_i\text{\bf -charges} & k_i\text{\bf -charges} 
\\\hline\hline 
(\underline{\sm \sfrac 12^3,\sfrac 12^2}, \sm \sfrac 12, ~\,\sfrac 12, ~\,\sfrac 12) &  
- \sfrac {a_i}2 - \sfrac {b_i}2 + \sfrac {c_i}2 + \sfrac {d_i}2  & k^{(1)}_i 
\\ 
(\underline{\sm \sfrac 12^3,\sfrac 12^2}, ~\,\sfrac 12, \sm\sfrac 12, ~\,\sfrac 12) &  
- \sfrac {a_i}2 + \sfrac {b_i}2 - \sfrac {c_i}2 + \sfrac {d_i}2  & k^{(2)}_i 
\\ 
(\underline{\sm \sfrac 12^3,\sfrac 12^2},  ~\,\sfrac 12, ~\,\sfrac 12, \sm\sfrac 12) &  
- \sfrac {a_i}2 + \sfrac {b_i}2 + \sfrac {c_i}2 - \sfrac {d_i}2  & k^{(3)}_i 
\\ 
(\underline{\sm \sfrac 12^3,\sfrac 12^2}, \sm \sfrac 12, \sm\sfrac 12, \sm\sfrac 12) &  
- \sfrac {a_i}2 - \sfrac {b_i}2 - \sfrac {c_i}2 - \sfrac {d_i}2  & k^{(4)}_i 
\\ 
(\underline{1^2,0^3}, 0^3)
& 2\, a_i & k^{(5)}_i
\\\hline 
}\nonumber
}
Note that this identification automatically builds in the constraint on the sum, $\sum_a k^{(a)}_i=0$. Solving for the quantities in \eqref{SU5inE8emb} we find
\equ{
\arry{lcl}{
a_i = - \sfrac 12\Big( ~\, k_i^{(1)} +k_i^{(2)} +k_i^{(3)} +k_i^{(4)} \Big)~, 
&\qquad &
b_i = - \sfrac 12\Big( ~\, k_i^{(1)} - k_i^{(2)} -k_i^{(3)} +k_i^{(4)} \Big)~, 
\\[1ex] 
c_i = - \sfrac 12\Big( \sm k_i^{(1)} +k_i^{(2)} -k_i^{(3)} +k_i^{(4)} \Big)~, 
&\qquad &
d_i = - \sfrac 12\Big(\sm k_i^{(1)} - k_i^{(2)} +k_i^{(3)} +k_i^{(4)} \Big)~. 
}
}
This shows in particular that when the sum of $k^{(1)}_i+\ldots+k^{(4)}_i$ is even (odd), we obtain the vectorial (spinorial) E$_8$ weights. These identifications should be read modulo permutations.

\subsection[A line bundle vector representation of \texorpdfstring{S(U(1)$^6$)}{S(U(1)\^{}6)} bundles]{A line bundle vector representation of \texorpdfstring{S(U(1)$^{\boldsymbol{6}}$)}{S(U(1)\^{}6)} bundles}

Similarly we can identify other line bundle backgrounds in both descriptions. As an example we consider S(U(1)$^6$) bundles with five independent bundle entries $k_{(a)}^i$, $a=1,\ldots, 5$.  In this case the unbroken gauge group is generically SU(4), hence we consider the branching 
\equ{ 
\rep{248} \ra (\rep{15},\rep{1}) + (\rep{1},\rep{45}) + (\rep{4},\rep{16}) 
+ (\crep{4},\crep{16}) + (\rep{6},\rep{10}) 
}
of $\text{E}_8 \ra \text{SU}(4) \times \text{SO}(10)$. The anti-symmetric tensor $\rep{6}$ can now be used to identify the translation uniquely. It is paired with the $\rep{10}$-plet, which branches as $\rep{10} \ra \rep{5} + \crep{5}$ under $\text{SO}(10)\ra \text{U}(5)$. Hence the five entries $k_{(a)}^i$ can be identified with the U(1) charges of the five components of the $\rep{5}$-plet in this branching. To realize this using the bundle vectors employed in this work, we take (up to overall permutations)
\equ{ \label{SU6inE8emb} 
V_i = (a_i^4, b_i, c_i, d_i,e_i)~. 
}
Matching the $\rep{6}$-plet charges, we find: 
\equ{ 
\arry{|c|c|c|}{ 
\hline 
\mathbf{6}\text{\bf -plets} & V_i\text{\bf -charges} & k_i\text{\bf-charges} 
\\\hline\hline  
(\underline{\sm \sfrac 12^2,\sfrac 12^2}, ~\, \sfrac 12, ~\, \sfrac 12, ~\,\sfrac 12, ~\,\sfrac 12) &  
~\, \sfrac {b_i}2 + \sfrac {c_i}2 + \sfrac {d_i}2 + \sfrac {e_i}2  & k^{(1)}_i 
\\ 
(\underline{\sm \sfrac 12^2,\sfrac 12^2}, \sm \sfrac 12, \sm \sfrac 12, ~\,\sfrac 12, ~\,\sfrac 12) &  
- \sfrac {b_i}2 - \sfrac {c_i}2 + \sfrac {d_i}2 + \sfrac {e_i}2  & k^{(2)}_i 
\\ 
(\underline{\sm \sfrac 12^2,\sfrac 12^2},  \sm \sfrac 12, ~\,\sfrac 12, \sm\sfrac 12, ~\, \sfrac 12) &  
- \sfrac {b_i}2 + \sfrac {c_i}2 - \sfrac {d_i}2 + \sfrac {e_i}2  & k^{(3)}_i 
\\ 
(\underline{\sm \sfrac 12^2,\sfrac 12^2}, \sm \sfrac 12, ~\, \sfrac 12, ~\,\sfrac 12, \sm\sfrac 12) &  
- \sfrac {b_i}2 + \sfrac {c_i}2 + \sfrac {d_i}2 - \sfrac {e_i}2  & k^{(4)}_i 
\\ 
(\underline{\sm 1^2,0^2}, 0^4)
& - 2\, a_i & k^{(5)}_i
\\\hline 
}\nonumber
}

\end{appendices}

{\small

\providecommand{\href}[2]{#2}\begingroup\raggedright\endgroup

}
\end{document}